\documentclass[conference,compsoc]{IEEEtran}
\usepackage{comment}

\ifCLASSOPTIONcompsoc
  \usepackage{cite}
\else
  \usepackage{cite}
\fi
\ifCLASSINFOpdf
\else
\fi
\usepackage{xspace}
\usepackage[shortlabels]{enumitem}
\usepackage[ruled,vlined,linesnumbered]{algorithm2e}

\SetCommentSty{mycommfont}
\usepackage{booktabs}
\usepackage{url}

\usepackage{xcolor}
\usepackage{listings}
\usepackage{multirow}
\usepackage{subfigure}
\usepackage{pifont}
\usepackage{threeparttable}

\usepackage[utf8]{inputenc}
\usepackage{array}
\usepackage{arydshln} 
\usepackage{color}
\definecolor{light-gray}{gray}{0.80}
\usepackage{mdframed}
\usepackage{framed}
\usepackage{cite}

\usepackage{float}
\usepackage{placeins}
\usepackage{multicol}

\newcommand{\ourSol}{\textsc{Cottontail}\xspace}

\newcommand{\ourSolNoValidator}{\textsc{Cottontail($\lnot$Val)}\xspace}

\newcommand{\ourSolNoSeedGen}{\textsc{Cottontail(Init+$\lnot$SGen)}\xspace}
\newcommand{\ourSolSeedGen}{\textsc{Cottontail(Init+SGen)}\xspace}

\newcommand{\ourSolRandomSeed}{\textsc{Cottontail(RandSeed)}\xspace}

\newcommand{\ourSolNormPro}{\textsc{Cottontail(NormPro)}\xspace}

\newcommand{\ourSoldp}{\textsc{Cottontail(dp-v3)}\xspace}
\newcommand{\ourSolnano}{\textsc{Cottontail(gpt-4.1-nano)}\xspace}
\newcommand{\ourSolmini}{\textsc{Cottontail(gpt-4o-mini)}\xspace}

\newcommand{\symcc}{\textsc{SymCC}\xspace}
\newcommand{\symccNoMap}{\textsc{SymCC($\neg$map)}\xspace}
\newcommand{\marco}{\textsc{Marco}\xspace}
\newcommand{\marcoMC}{\textsc{Marco(mc)}\xspace}
\newcommand{\marcoCFG}{\textsc{Marco(cfg)}\xspace}

\newcommand{\tightcolorbox}[2]{%
  \begingroup
  \setlength{\fboxsep}{2pt}% no padding
  \colorbox{#1}{#2}%
  \endgroup
}

\colorlet{punct}{red!60!black}
\definecolor{background}{HTML}{EEEEEE}
\definecolor{delim}{RGB}{20,105,176}
\colorlet{numb}{magenta!60!black}

\lstdefinelanguage{json}{
    basicstyle=\small\ttfamily,
    numbers=none,
    numberstyle=\scriptsize,
    stepnumber=1,
    numbersep=5pt,
    showstringspaces=false,
    breaklines=true,
    frame=lines,
    backgroundcolor=\color{background},
    literate=
     *{0}{{{\color{numb}0}}}{1}
      {1}{{{\color{numb}1}}}{1}
      {2}{{{\color{numb}2}}}{1}
      {3}{{{\color{numb}3}}}{1}
      {4}{{{\color{numb}4}}}{1}
      {5}{{{\color{numb}5}}}{1}
      {6}{{{\color{numb}6}}}{1}
      {7}{{{\color{numb}7}}}{1}
      {8}{{{\color{numb}8}}}{1}
      {9}{{{\color{numb}9}}}{1}
      {:}{{{\color{punct}{:}}}}{1}
      {,}{{{\color{punct}{,}}}}{1}
      {\{}{{{\color{delim}{\{}}}}{1}
      {\}}{{{\color{delim}{\}}}}}{1}
      {[}{{{\color{delim}{[}}}}{1}
      {]}{{{\color{delim}{]}}}}{1},
}
\usepackage[T1]{fontenc}
\usepackage{textcomp}
\usepackage{tikz}
\usepackage{hyperref}
\hypersetup{
    colorlinks=true,
    linkcolor=blue,
    filecolor=magenta,      
    urlcolor=black,
    citecolor=blue,
    pdftitle={Overleaf Example},
    pdfpagemode=FullScreen,
}

\definecolor{mygreen}{rgb}{0,0.5,0}
\definecolor{myblue}{rgb}{0,0,1}
\definecolor{mymauve}{rgb}{0.58,0,0.82}
\definecolor{myblack}{rgb}{0.24,0.17,0.12}
\definecolor{awesome}{rgb}{1.0, 0.13, 0.32}

\begin{document}
\title{\ourSol: Large Language Model-Driven Concolic Execution for \\ Highly Structured Test Input Generation}

\author{\IEEEauthorblockN{Anonymous Authors}}


\author{
  
\IEEEauthorblockN{Haoxin Tu}
\IEEEauthorblockA{Singapore Management University \\
haoxintu@gmail.com}
\\
\IEEEauthorblockN{Peng Chen}
\IEEEauthorblockA{Independent Researcher\\
spinpx@gmail.com}
\and

\IEEEauthorblockN{Seongmin Lee}
\IEEEauthorblockA{University of California, Los Angeles\\
seongminlee@sigsoft.org}
\\
\IEEEauthorblockN{Lingxiao Jiang}
\IEEEauthorblockA{Singapore Management University\\
lxjiang@smu.edu.sg}

\and
\IEEEauthorblockN{Yuxian Li}
\IEEEauthorblockA{Singapore Management University\\
liyuxianjnu@gmail.com}
\\
\IEEEauthorblockN{Marcel Böhme}
\IEEEauthorblockA{Max Planck Institute for Security and Privacy\\
marcel.boehme@acm.org}
}


% compsoc conferences compared to traditional conferences), use this
%Telephone: (800) 555--1212, Fax: (888) 555--1212}


\maketitle
\pagestyle{plain}
\thispagestyle{plain}
\begin{abstract}
 How can we perform concolic execution to generate highly structured test inputs for systematically testing parsing programs? 
 Existing concolic execution engines are significantly restricted by 
 (1) input structure-agnostic path constraint selection, leading to the waste of testing effort or missing coverage; 
 (2) limited constraint-solving capability, yielding many syntactically invalid test inputs;
 (3) reliance on manual acquisition of highly-structured seeds, resulting in non-continuous testing.

 This paper proposes \ourSol, a new Large Language Model (LLM)-driven concolic execution engine, to mitigate the above limitations.
 A more complete program path representation, named Expressive Coverage Tree (ECT), is first constructed to help select structure-aware path constraints.
 Later, an LLM-driven constraint solver based on a {\it Solve-Complete} paradigm is designed to solve the path constraints smartly to get test inputs that are not only satisfiable to the constraints but also valid to the input syntax.
 Finally, a history-guided seed acquisition is employed to obtain new highly structured test inputs either before testing starts or after testing is saturated.
 We implemented \ourSol on top of \symcc and evaluated eight extensively tested open-source libraries across 
 four different formats (XML, SQL, JavaScript, and JSON). 
 The experimental results are promising: \ourSol significantly outperforms baseline approaches by 30.73\% and 41.32\% on average in terms of line and branch coverage.
 Besides, \ourSol found six previously unknown vulnerabilities (six CVEs assigned). We have reported these issues to developers, and four out of them have been fixed so far.
\end{abstract}


% \fi
\IEEEpeerreviewmaketitle

\newsavebox{\mybox}

\section{Introduction}  \label{cottontail::sec:introduction}

%\hlt{Background.} 
Parsing software systems, such as XML and SQL libraries, are widely used in modern systems. 
However, even after years of intensive testing efforts, residual vulnerabilities persist, reflecting the complexity and attack surface of such components.
Highly structured (or syntactically valid) test inputs are demanded to comprehensively stress the parsing test programs; as only the parser-checking logic is passed, the deeper application logic can be examined. 
Considerable effort has been devoted to generating structured test inputs, including black, grey, and white-box fuzzing-based approaches. 
Among them, white-box fuzzing via concolic execution has shown considerable capabilities of test input generation for general test programs. 
Given a seed input, a concolic execution engine starts by concretely executing the program while symbolically tracking the same execution path to collect path constraints. 
The negation of path constraints is applied to explore alternative branches. 
An off-the-shelf constraint solver is used to solve constraints and generate new test cases
that satisfy the negated constraints, enabling the path exploration of uncovered paths. 
Benefiting from the soundness of test case generation and a systematic way for path exploration, it has been promising and applied in many areas \cite{SymCC-usenix20,SymSan-usenix22,hu2024marco,yun2018qsym-usenix18,boosting-se,krover,fastklee}. 

%\hlt{Problem.} 
Although promising, existing concolic executors (e.g., \symcc \cite{SymCC-usenix20} and \marco \cite{hu2024marco}) remain significantly hindered by three fundamental limitations in their treatments for the problems of {\it which to solve}, {\it how to solve}, and {\it how to acquire new seed inputs} when handling parsing test programs.

\smallskip
\noindent
{\bf \#L1: The input structure-agnostic path constraints selection is either redundant or overly aggressive}.
When path constraints are generated during concolic execution, we need to consider the problem of {\it which path constraints to solve}.
A straightforward idea is to select all path constraints, hoping to explore all paths in test programs. However, such a structure-agnostic option leads to many redundant path constraints, making the testing process impractical.
Alternatives are to select the path constraints based on some heuristics, e.g., bit-wise coverage map {\tt Bitmap} from \symcc \cite{SymCC-usenix20} and Concolic State Transition Graph (CSTG) from \marco \cite{hu2024marco}.
Unfortunately, both guides are built on the basis of binary code, which aggressively eliminates interesting code coverage. This is mainly due to the lack of expressive coverage information in the path constraints, making them difficult to distinguish and select interesting paths (detailed in \S\ref{cottontail::sec:approach:why-esct}).

%\item 
\smallskip
\noindent
{\bf \#L2: The solutions from constraint solving only comply with satisfiability while neglecting syntactic validity.}
To obtain high-quality (especially for highly structured) test cases by solving constraints, an important question is  {\it how to solve the constraints}. 
Traditional constraint solvers (e.g., {\tt Z3}) equipped with concolic executors usually solve constraints for satisfiability, while ignoring the syntactic validity of newly generated test cases (see more explanation in \S\ref{cottontail::sec:background:investigation}). 
Such a limitation oftentimes causes the engine to produce syntactically invalid test cases, rendering the testing effort largely in vain. 
We argue that an optimal solution for constraints should not only comply with satisfiability, but also be aware of syntactic validity.
Also, traditional solvers can not generate new test cases with flexible sizes by design, as the number of symbolized bytes is restricted by the seed, further decreasing the effectiveness of concolic testing.

%\item 
\smallskip
\noindent
{\bf \#L3: The acquisition of highly structured seeds before testing starts or after testing is saturated is difficult.}
Existing engines highly rely on manually collected {\it initial} seeds from bug repositories to start testing, where the manual work is often time-consuming.
%Using LLMs to 
Randomly generating seeds could be an alternative, but it tends to be ineffective, as many random seeds do not boost coverage, and it would be wasteful to continue testing if the testing is saturated (i.e., code coverage has plateaued). 
We thus need to acquire {\it fresh} seeds to change the situation, but existing concolic executors are unable to generate such seed inputs during testing progress.
Again, it is possible to naively feed random seeds into the testing process after saturation, but it rarely improves code coverage (as demonstrated in \S\ref{cottontail::eva:rq2.4-seed-generation}).
%\end{itemize}

\smallskip
To overcome the above limitations, we propose \ourSol\footnote{Cottontail rabbits are known for their structured running patterns (e.g., zigzagging) to evade predators using their cotton-ball tails. 
We used it to reflect our aim for the generation of highly structured test input.}, a new Large Language Model (LLM)-driven concolic execution engine to generate highly structured test inputs effectively. 
Our key insights are threefold. 
{\it First}, 
We found that for parsing programs, the inputs are processed structurally, e.g., using structured branches (either {\it switch-case}, {\it if-else}, or others) to handle different input bytes.
Thus, we can build what we call \emph{structural program paths}, i.e., paths that diverge depending on specific input values, to help not only represent meaningful and complete program paths but also reduce redundant path constraints, addressing limitation \textbf{\#L1}.
{\it Second}, given the strong input understanding and completion capabilities of LLMs \cite{lu2021codexglue}, it could be promising to leverage them to (1) perform syntax-aware solving, i.e., {\it solve} the constraints for not only satisfiability but also syntax validity; (2) conduct flexible solution completing, i.e., {\it complete} the solution to be syntactically valid and with flexible sizes, thus alleviating limitation \textbf{\#L2} (more details in \S\ref{cottontail::sec:background:investigation}).
{\it Third}, benefiting from LLM knowledge and memorization \cite{nam2024using, sun2024source, ma2023lms,LLMCodeAnalysis-usenix24}, LLM could be induced to produce new meaningful seed inputs, thus mitigating limitation \textbf{\#L3}.

%\hlt{Our approach.}
Based on the above three insights, we design three new components in \ourSol to explore structural program paths for highly structured test input generation.

\begin{itemize}[leftmargin=1em,nosep]
\item {\bf Structure-aware Constraint Selection}. 
To address the limitation of {\it which to solve}, a branch information collector is first introduced during the instrumentation to help construct a more complete representation of {\it structural program path}. Then, based on the representation, a new coverage map, named Expressive Coverage Tree (ECT), is constructed to keep track of program branch status (e.g., taken or untaken) with expressive semantic information. 
Finally, guided by ECT, a path constraint selector is facilitated to conduct structure-aware path selection.

\item {\bf LLM-driven Constraint Solving}. To mitigate the limitation of {\it how to solve}, an LLM-driven solver, which facilitates {\it Solve-Complete} paradigm, is leveraged to solve the path constraints.
The paradigm first {\it solves} constraints for satisfiability and then {\it completes} the solution for the syntax validity.
Also, the LLM-generated test cases can have a flexible size, alleviating the restrictions of generating only test cases with fixed sizes.
Moreover, to increase the robustness of LLMs, a test case validator is designed to refine unsound results from LLMs. 

\item {\bf History-guided Seed Acquisition}. To address the limitation of {\it how to acquire new seeds}, history-guided seed acquisition strategies are designed for different timings. 
Before testing, we prompt LLMs to generate highly structured test inputs as {\it initial} seeds that may likely trigger new vulnerabilities based on their memories (e.g., from historical bug repositories).
During testing, a history coverage recorder is utilized to record the mapping between the test input and its code coverage.
When the coverage gets saturated, we prompt LLMs with Chain-of-Thought (CoT) \cite{cot-llms} based on the historical mapping to generate {\it fresh} seeds that are likely to cover unexplored features.

\end{itemize}

%\smallskip
In short, \ourSol is a novel concolic execution engine that is capable of {\it structure-aware constraint selection}, {\it smart constraint solving}, {\it history-guided seed acquisition} for robust generation of highly structured test inputs.
We position \ourSol as a new white-box fuzzing that can effectively explore both {\it input space} 
and {\it execution space} to advance the field of automated testing (see more detailed comparison with existing structure-aware fuzzing in \S\ref{cottontail::sec:related-work}).

%\hlt{Evaluation.}
We have prototyped \ourSol on top of \symcc \cite{SymCC-usenix20} and demonstrated its test input generation capabilities over eight widely tested libraries across four different formats ({\tt XML}, {\tt SQL}, {\tt JavaScript}, and {\tt JSON}).
Our experiments show promising results.
Compared with state-of-the-art approaches (i.e., \symcc \cite{SymCC-usenix20}, \marco \cite{hu2024marco}, and their variants), \ourSol significantly outperforms them by covering 30.73\% more lines and 41.32\% more branches on average.
During the same period, \ourSol significantly improves (more than 100x) the parser checking passing rate over the generated test cases.
Our ablation studies also demonstrate that each component in \ourSol has contributed to the better results.
We have also found six previously unknown memory-related vulnerabilities and reported them to the developers (six new CVE IDs have been assigned to them, and four out of six have been fixed).

\smallskip
{\bf  \textit{Contributions.}} We make the following contributions:
\begin{itemize}[leftmargin=1em,nosep]
  \item To our knowledge, \ourSol is the first LLM-driven concolic execution engine for highly structured test input generation, automatically working in a white-box manner.
  \item Three new components, including structure-aware path constraint selection, smart LLM-driven constraint solving, and history-guided seed acquisition, are designed to make \ourSol effective and practical.
  \item Extensive experiments are conducted to demonstrate the capabilities of \ourSol. The results show that \ourSol can not only significantly outperform baseline approaches but also is practical to find new vulnerabilities.
  \item The prototype \ourSol is open source\footnote{%Detailed description of the artifact installation and usage can be found in Section \ref{cottontail::appendix:artifact} in the Appendix).
\url{https://github.com/Cottontail-Proj/cottontail}} to foster future research that combines program analysis and LLMs.
\end{itemize}

%{\bf Organizations.} 
%Section \ref{cottontail::sec:motivation} gives the background and motivation. 
%Section \ref{cottontail::sec:approach} describes the design of \ourSol. 
%Section \ref{cottontail::sec:evaluation} presents the implementation of \ourSol and our evaluation results. 
%Section \ref{cottontail::sec:discussion} discusses the cost of our experiments, threats to validity, limitations, and others.
%Sections \ref{cottontail::sec:related-work} and \ref{cottontail::sec:conclusion} describe related works and conclude with future work. 

\section{Background and Motivation}  \label{cottontail::sec:background}

\begin{figure}[t]
	\centering
	\includegraphics[width=0.90\linewidth]{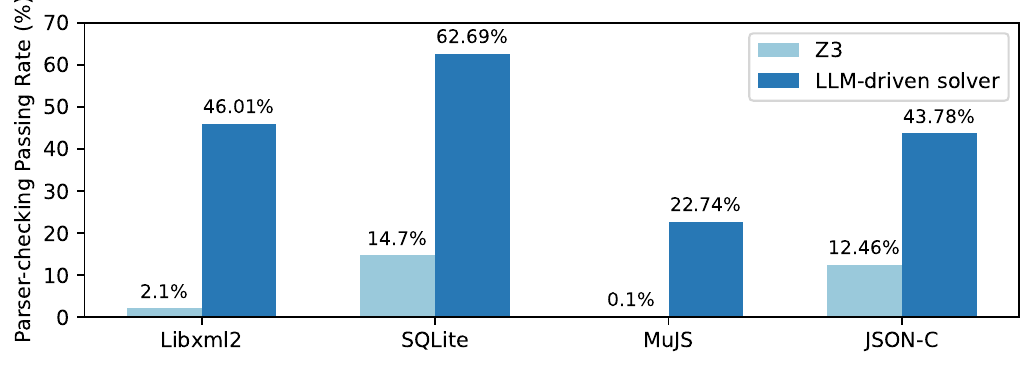}
	\vspace{-0.5em}
	\caption{Comparison of parser checking pass rates between traditional solver (i.e., {\tt Z3}) and LLM-driven solver (designed in \ourSol). }
	\label{cottontail::fig:llm-solving-intro}
	\vspace{-1em}
\end{figure}

\noindent
{\bf Concolic Execution.}
Concolic execution, also known as dynamic symbolic execution, integrates symbolic and concrete execution to explore program paths systematically. 
Concrete values guide the actual execution path, ensuring feasibility, while symbolic values enable exploration of alternative paths by generating new test cases. In recent years, compilation-based concolic execution (e.g., \symcc \cite{SymCC-usenix20} and \marco \cite{hu2024marco}) has gained popularity due to its superior performance and practical applicability.
In particular, concolic execution is a key component of the tools for the winning team for both old DARPA CGC \cite{CGC,mayhem} and recent AIxCC Atlanta \cite{AIxCC}.
Technically, given an initial seed input, these engines embed symbolic reasoning/tracing logic directly into compiled binaries and collect path constraints at runtime. 
By negating selected path constraints, the engine produces new constraints representing unexplored branches, which are then solved using off-the-shelf constraint solvers (e.g., {\tt Z3} \cite{z3}) to generate new test cases.
Note that the new test cases usually hold the same size as the seed input because the size of symbolic bytes is fixed when the seed input is fed into the concolic engine in the initial phase.
The concolic testing process is continued by iteratively feeding the new test cases back into the execution loop, which is ideally an endless process. 
However, concolic testing naturally encounters a saturation point when it can no longer cover new code due to the lack of diversity in test inputs (e.g., limited by the input size) \cite{mayhem,symsize,SymLoc} or the restricted covering capability of harnesses/test drivers \cite{promptfuzz,FuzzGen,UTopia}. 
When testing is saturated, the best practice is to acquire {\it fresh} new seeds that drive the exploration forward for long-term continuous testing.
In short, {\it which to solve} determines both effectiveness and efficiency of testing, {\it how to solve} affects the effectiveness of test case generation, while {\it how to acquire new seeds} decides the continuity of testing.

%\subsection{LLMs for Test Input Generation}  \label{cottontail::sec:background:LLMs}
\smallskip
\noindent
{\bf LLMs for Test Input Generation.}
Large Language Models (LLMs) are popular AI systems designed to predict the next word or token in a sequence based on the context of preceding tokens. 
Recent emergence of LLMs has driven their application in numerous security-related domains \cite{lin2025large,pearce2023examining,concollmic}. 
However, most of the LLM-based systems lack analytical depth and robustness, which limits their effectiveness in more complex scenarios such as systematic program understandings~\cite{LLMEvaSP24}.
Two promising directions to mitigate the problem are either to integrate advanced program analysis or design logical Chain-of-Thought (CoT) prompts to further improve the reliability and robustness of the LLM-based system \cite{llmSurvey,nong2024chainofthoughtpromptinglargelanguage}.
While LLMs have been combined with black-box \cite{xia2024fuzz4all,titanFuzz} or grey-box \cite{chatafl-ndss24,promptfuzz} fuzzing techniques for structured test input generation, to our knowledge, no study has yet attempted to integrate LLMs with more systematic techniques like concolic execution to enhance the security analysis. 
We believe it could be promising to empower the potential of LLMs for more rigorous security guarantees by integrating LLMs with concolic testing.

%\subsection{Investigation Study} \label{cottontail::sec:background:investigation}

\begin{figure}[t]
	\centering
	\includegraphics[width=0.98\linewidth]{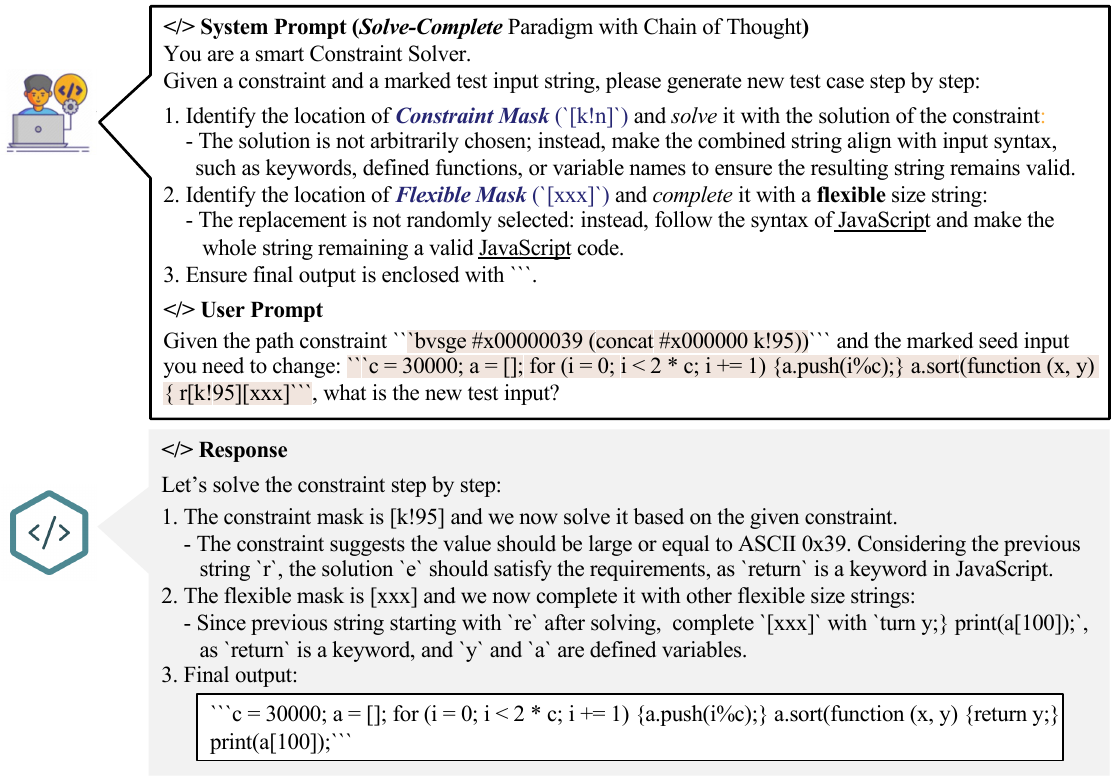}
	%\vspace{-1em}
	\caption{The {\it Solve-Complete} paradigm and LLM's response. In the upper box, the underlined \underline{text} represents different formats, and the colored \colorbox{light-gray}{\makebox(10,4){text}} enclosed %(with  \textasciigrave \textasciigrave \textasciigrave) 
    indicates either constraints or a marked test input string.}
	\label{cottontail::fig:cot-solving}
	\vspace{-1em}
\end{figure}

\begin{figure*}[t]
	\centering
	\includegraphics[width=0.98\linewidth]{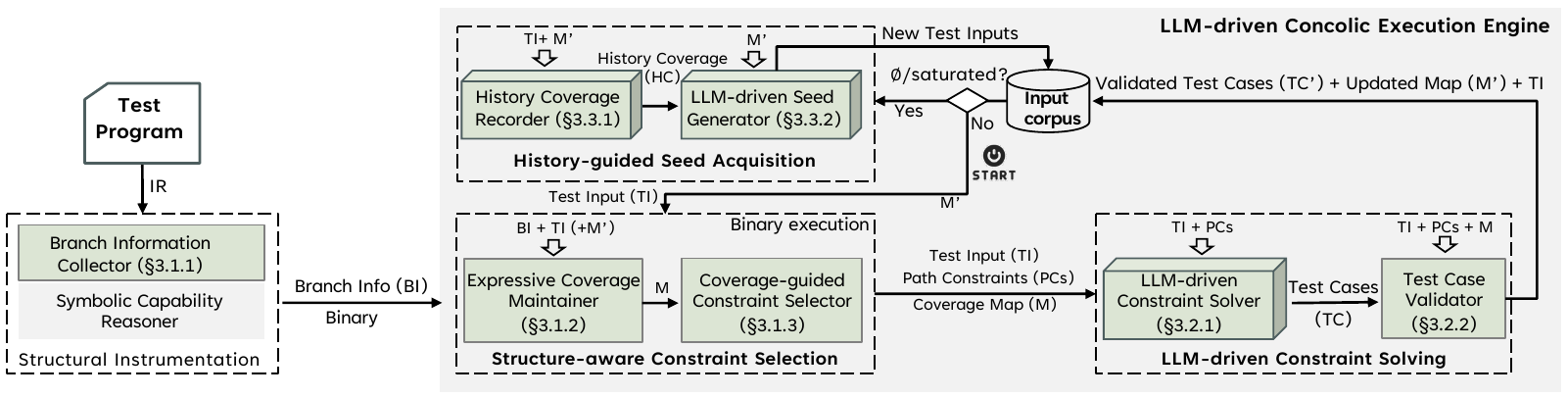}
	%\vspace{-0.5em}
	\caption{High-level design of \ourSol concolic execution engine}
	\label{cottontail::fig:overview}
	%\vspace{-1em}
\end{figure*}

\smallskip \label{cottontail::sec:background:investigation}
\noindent 
{\bf Motivation: Investigative Study.}
It is evident that there is a need for a new path constraint selection strategy to select optimal subsets of constraints and a new seed acquisition to generate highly structured inputs to improve both effectiveness and continuity of concolic testing. 
It may not be clear why we need an LLM-driven solver (which served as our core innovation) to generate highly structured test inputs.

To justify our motivation, we conducted an investigation study to show the significant limitations of traditional constraint solvers and how an LLM-driven constraint solver can mitigate them.
To do so, we first selected four libraries that process four different input formats ({\tt XML/SQL/JavaScript/JSON}) as test programs, and ran them using the existing concolic executor \symcc.
Then, we collected the generated path constraints and used a traditional solver {\tt Z3} to solve them to obtain new test cases.
After running the new test cases with four target programs, we found that numerous test cases (85.3\% to 99.9\%, as shown in Figure~\ref{cottontail::fig:llm-solving-intro}) are syntactically invalid and cannot pass the parser checking logic.
This is reasonable as existing solvers only solve constraints for satisfiability while ignoring the guarantee of syntactic validity of the solution by design. 
Also, the size of the resulting solution is fixed, restricting the diversity of generated test cases.
For example, given a path constraint ``{\it bvsge \#x00000039 (concat \#x000000 k!95)}'' (where `k!95' is a symbolic variable, collected from parsing implementation in Figure~\ref{cottontail::fig:mujs-parsing}) which requires the value of the symbol `k!95' to be large or equal to ASCII 0x39 (i.e., char `{\it 9}'). 
{\tt Z3} simply solves it to `{\it 9}' and keeps the remaining string unchanged with fixed size as seed input, producing an invalid JavaScript string `{\it r9turn ...}'. 
This fact indicates that the test cases generated by existing concolic executors can hardly examine the deeper code regions (e.g., the application logic), significantly hindering the effectiveness of concolic testing.
In summary, such a limitation motivates us to investigate the following question: {\it How to solve the constraints smartly, i.e., solving constraints for both satisfiability and syntactic validity as well as making the resulting solutions with flexible sizes}?

To answer the question, we design a new LLM-driven constraint solver based on {\it Solve-Complete} paradigm as shown in  Figure~\ref{cottontail::fig:cot-solving}.
The idea behind it is simple but effective. 
Given a path constraint and the marked (two types of marks help {\it Solve-Complete} paradigm) seed input string as a user prompt, LLMs are prompted to smartly solve the given constraint within two consecutive steps.
In step (1), LLMs are asked to {\it solve} the constraint and use the solution to replace the {\it Constraint Mask} `[k!n]', where the solution is syntax-aware, that is, supposed not only to meet the satisfiability but also possibly comply with the input syntax validity. 
In step (2), LLMs are required to {\it complete} the {\it Flexible Mask} `[xxx]' with a flexible size string which makes the whole resulting string remain valid.
Taking the same path constraint includes `k!95' again, our solver solves it to `{\it e}', which can be connected with a remaining string (e.g., `{\it turn}') to form a syntactically valid JavaScript string (as shown in the LLM's response).
To demonstrate the effectiveness of the LLM-driven solver, we used it to solve the same path constraints solved by {\tt Z3}, and the results presented in Figure~\ref{cottontail::fig:llm-solving-intro} show significant improvements (100x more) in terms of the parser checking pass rate across various formats, practically helping achieve significant improvement in terms of code coverage.

\section{Design of \ourSol}  \label{cottontail::sec:approach}

\smallskip
{\bf Overview.} 
Figure~\ref{cottontail::fig:overview} presents the high-level design of \ourSol. 
Given a test program, \ourSol enhances instrumentation by collecting branch information at compile-time and constructs a new program path representation called Expressive Coverage Tree (ECT) at runtime to assist in structure-aware constraint selection (\S\ref{cottontail::sec:approach:constraint-selection}).
Later, \ourSol leverages LLM-driven constraint solving with {\it Solve-Complete} paradigm to solve constraints smartly and refine the non-robust results from LLMs (\S\ref{cottontail::sec:approach:constraint-solving}).
Finally, if there are no initial seed inputs to set up testing or whenever the testing process gets saturated, \ourSol adopts a history-guided seed acquisition to obtain high-quality seed inputs continuously (\S\ref{cottontail::sec:approach:seed-generation}).

\subsection{Structure-aware Constraint Selection} \label{cottontail::sec:approach:constraint-selection}

This subsection explains why a new path constraint selection strategy is needed and details how we construct a new coverage map to guide the structure-aware selection.

\subsubsection{Why Structure-aware Selection}  \label{cottontail::sec:approach:why-esct}
To perform a practical concolic execution over parsing test programs, we argue that an effective selection should be structure-aware, which: %\seongmin{feel free to rollback this refactoring} The selection strategy
\begin{enumerate}[label=\#\arabic*]
    \item provides a meaningful (e.g., includes semantic information) and complete representation of program paths;
    \item records human-readable coverage information;
    \item excludes redundant structure-agnostic path constraints;
    \item has less chance of missing interesting coverage.
\end{enumerate}

Satisfying requirement \#1 helps users to have a better understanding of structural paths, \#2 is essential to help craft useful inputs either by humans or LLMs at runtime when the testing gets saturated. Compared with branch information from binary code (which contains less semantic information), a human-readable coverage could provide clear and interpretable insights into which branches have been covered and what direction to create new test cases. \#3 and \#4 together guarantee a better trade-off between testing effectiveness and efficiency.
There are three existing strategies to handle the path constraint selection for the general software systems, which are exhaustive search, the Bitmap-based approach \cite{SymCC-usenix20}, and the CSTG-based approach \cite{hu2024marco}, yet they do not comply with all the requirements when handling parsing test programs.
Justified by the above facts, it thus calls for a new path constraint selection that could satisfy all four requirements.
In the following subsections, we detail the structural instrumentation and human-readable Expressive Coverage Tree (ECT) to meet requirements \#1 and \#2, and the ECT-guided path constraint selector to comply with requirements \#3 and \#4.

\begin{figure}[t]
\centering

\begin{lrbox}{\mybox}%
\begin{lstlisting}[escapechar=@, frame=single] 
// Parsing logic /* jslex.c */
static int jsY_isidentifierpart(int c) {
    return isdigit(c) //``bvsge #x00000039 (concat #x000000 k!95)''
    || isalpha(c) || c == '$' ||c == '_' ||isalpharune(c);
} 
static int jsY_lexx(js_State *J){
    while (1) {
        //...
        switch (J->lexchar) {
	       case '(': jsY_next(J); return '(';
		   case ')': jsY_next(J); return ')';
		   case ',': jsY_next(J); return ',';
            //...
        }
	//...
    }
} 
\end{lstlisting} 
%\vspace{-1em}
\end{lrbox}%
\scalebox{0.9}{\usebox{\mybox}}
\caption{Sample parsing implementation code from MuJS}\label{cottontail::fig:mujs-parsing}
%\vspace{-1em}
\end{figure}

\subsubsection{Structural Instrumentation}\label{cottontail::sec:approach:instrumentation}

Compilation-based concolic execution has shown promising performance in execution speed compared with IR (Intermediate Representation)-based execution, but it inevitably misses many interesting behaviors of the test programs due to the loss of semantics information after compilation \cite{SymCC-usenix20,SymSan-usenix22}. 
To alleviate this issue, we propose to use extra instrumentation on the IR code to collect necessary branch information.
It is worth noting that such instrumentation is crucial for capturing meaningful structural information, particularly around complex branch conditional constraints.
This is because it allows developers and analysis tools to observe precisely which paths of the code are being exercised at runtime. 
Without such instrumentation, it can be difficult to determine whether certain cases or branches within a switch statement are triggered, leading to potential blind spots in testing. 
To comprehensively cover structural program paths, we design a branch information collector to capture all possible structural paths during instrumentation.
To be specific, our structural instrumentation systematically walks through every instruction in a given function, looking for branch instructions such as {\it switch}. When it encounters a branch, it records the branch name and its associated case values. Such metadata is then added to a global map that associates each branch statement with its case values. Finally, the instrumentation phase saves all gathered information into a {\tt JSON} file, enabling further analysis of the function's branching structures and program semantics.

\subsubsection{Expressive Coverage Tree Maintainer}  \label{cottontail::sec:approach:structural-coverage-maintainer}

After collecting branch information, we introduce a new Expressive Coverage Tree (ECT) to help have a comprehensive representation of structural program paths.

We define ECT as a hierarchical tree structure represented by a pair \( T = (N, E) \), where:
\begin{itemize}[leftmargin=1em,itemsep=0.1cm]
    \item \( N \) is a set of nodes, containing a special root node. 
    \item \( E \subseteq N \times N \) is a set of edges that define parent-child relationships, representing the call context (caller to callee) or branch information (condition to nested statement).
\end{itemize}

Each node \( n \in N \) may have zero or multiple child nodes connected by edges. 
Nodes without children are called {\it leaf} nodes; nodes with children are referred to as {\it internal} nodes. 
Each node name is a unique identifier:\\
\centerline{{\it fileName\_funcName\_lineNum\_colNum\_brType\_brId}}
which consists of several important attributes to represent a unique branch or differentiate between different branches. 
Those attributes include visiting status ({\it taken} or {\it untaken}), visit count ({\it visit\_cnt}), branch type ({\it brType}), call stack size, and branch id ({\it brId}) --- in {\it if} statement, it refers to 0 ({\it then} branch) or 1 ({\it else} branch); in {\it switch-case} statement, it represents the constant case value.
With the help of the expressive coverage map, users can easily understand the testing process by checking the statistics in the global map.

\begin{figure}[h]
\centering
%\begin{lrbox}{\mybox}%
\begin{lstlisting}[escapechar=@, frame=single,numbers=none] 
{
"loc": "jslex.c_jsY_lexx_9_3_switch", "tp": 1, "tk": 1, "cs": 10, "vc": 1, "br": -1,
    "ch": [
	{ "loc": "jslex.c_jsY_lexx_9_3_switch_40", 
		  "tp": 1, "tk": 1, "cs": 10, "vc": 1, "br": 40 },
	{ "loc": "jslex.c_jsY_lexx_9_3_switch_41", 
		  "tp": 1, "tk": 1, "cs": 10, "vc": 1, "br": 41 },
	{ "loc": "jslex.c_jsY_lexx_9_3_switch_44", 
		  "tp": 1, "tk": 1, "cs": 10, "vc": 1, "br": 44 }
	]
}
\end{lstlisting} 
\caption{Partial expressive coverage tree (ECT) recording the program paths between Lines 9-12 in Figure \ref{cottontail::fig:mujs-parsing} in JSON format ({\it loc}: source code location; {\it ch}: children; {\it tp}: branch type (0 for {\it if} statement and 1 for {\it switch} statement); {\it tk}: taken; {\it cs}: call stack size; {\it vc}: visited count; {\it br}: branch id).}\label{cottontail::fig:ect-json}
\label{cottontail::fig:ect-json}
%\vspace{-1em}
\end{figure}

To provide a clearer understanding of the ECT used in \ourSol, 
Figure \ref{cottontail::fig:ect-json} presents a partial ECT that captures the coverage information of the switch statement with three children in Figure \ref{cottontail::fig:mujs-parsing} at Line 9.
Once the function {\tt jsY\_lexx} is analyzed, the ECT of the program in JSON format is recorded in the global coverage map.
It is worth noting that the ECT differs from existing code coverage trees (e.g., CSTG from Marco \cite{hu2024marco}) or maps (e.g., the bitmap from SymCC \cite{SymCC-usenix20}): It is \emph{partially} context-sensitive and path sensitive, capturing not only branch coverage information but also call stack information.
Furthermore, since our representation is based on source code instead of binary code, we can avoid the loss of interesting path coverage or semantic information (e.g., detailed {\it switch-case} branches to store structured information) due to hash collision or source code compilation.
In short, we provide a precise representation of code coverage that enables the distinction of different execution contexts for the same branch, thereby facilitating systematic exploration of the input space and advancing automated test generation.

After defining ECT, we manage and update a global coverage tree to help guide the constraint selection.
Thus, \ourSol is able to reduce redundant coverage but avoid losing promising code coverage.

\subsubsection{ECT-guided Constraint Selector}  \label{cottontail::sec:approach:constraint-selector}

The selector consists of two phases of reduction: reducing redundant constraints that do not increase coverage during single concolic execution, and reducing constraints across different runs that have less chance to boost code coverage.

\smallskip
\noindent
{\it First Phase Reduction.}
As aforementioned in Figure \ref{cottontail::fig:mujs-parsing}, in a single concolic execution, it is common that the redundant path constraints are collected when each input byte is repeatedly analyzed by a parsing function. 
Therefore, we need to remove such duplicated path constraints.
To do so, we maintain a global branch recorder to record branch-constraint mappings during a single concolic execution.
Since each element in the record is a unique branch identifier that records the context of the branch.
When a branch is encountered and already exists in the recorder, we compare the current constraints and the constraints stored in the branch.
If the constraints are only different in the symbolic index (i.e., each byte represented as `k!n' in path constraints, where `n' indicates the index over the input bytes), as shown in Figure \ref{cottontail::fig:reduction-phase1}, we reduce a duplicate set of path constraints and only keep unique branches.

Note that such a constraint deduplication mechanism preserves the soundness of concolic execution by excluding only redundant path constraints that arise from structurally identical branches applied repeatedly across input positions (also demonstrated in first phase selection in Table \ref{cottontail::tab:evaluation:rq2.2-selector}).  
%As a result, our approach reduces constraint redundancy without omitting meaningful paths, thus preserving the soundness of the analysis (also demonstrated in first phase selection in Table \ref{cottontail::tab:evaluation:rq2.2-selector}).

\begin{figure}[t] 
\centering
\begin{lrbox}{\mybox}%
\begin{lstlisting}[escapechar=@,frame=none]
(vsge #x00000039 (concat #x000000 k!0)
(vsge #x00000039 (concat #x000000 k!1))
....
(vsge #x00000039 (concat #x000000 k!95))

\end{lstlisting} 
\end{lrbox}%
\scalebox{1}{\usebox{\mybox}}
%\vspace{-1em}
\caption{Duplicated path constraints in parsing logic (those path constraints aim to cover the same branch at Line 3 in Figure \ref{cottontail::fig:mujs-parsing})
}
%\vspace{-1em}
\label{cottontail::fig:reduction-phase1}
\end{figure}

\smallskip
\noindent
{\it Second Phase Reduction.}
Different from the first phase, in the second phase, we remove redundant path constraints based on the newly built coverage tree across different runs.
An important problem to handle is how we remove path constraints without affecting the overall performance (i.e., missing potential interesting code coverage). 
Simply excluding the branches that have been explored tends to miss much interesting coverage, as such a strategy does not have a chance to examine the remaining execution of the current test case or the remaining test cases to make a globally optimal decision \cite{hu2024marco}.
Therefore, we need to be careful when making the selection decision.
Inspired by prior work, such as KLEE \cite{klee} and Mayhem \cite{mayhem,stephens2016driller}, we consider factors including untaken branches, frequency of visits, and depth of execution to balance exploration potential and redundancy, as each coefficient has been shown to contribute to uncovering previously unexplored paths.
Thus, we propose a new metric node weight to quantify selection priority, defined as follows:
\begin{equation} \label{cottontail::eq:selector}
\mathit{Node}_{weight} = \alpha \cdot untaken + 
\beta \cdot visit\_cnt + 
\gamma \cdot depth
\end{equation}

The {\it untaken}, {\it visit\_cnt}, and {depth} are node attributes, and $\alpha$, $\beta$, and $\gamma$ are three parameters to optimize exploration for maximum program coverage. 
$\alpha$ prioritizes untaken branches, ensuring the discovery of new execution paths and highlighting untested code regions. 
$\beta$ focuses on rarely visited nodes, balancing exploration by avoiding overemphasis on frequently traversed paths while paying attention to less common scenarios.
$\gamma$ rewards deeper nodes, encouraging exploration of complex execution paths and uncovering deeply nested bugs or vulnerabilities. 
Together, these parameters enable a balanced trade-off that drives efficient and thorough program testing.

%\subsubsection{Running Example}

%\todo{add something if time allows}

\subsection{Smart LLM-driven Constraint Solving} \label{cottontail::sec:approach:constraint-solving}

This subsection introduces our {\it Solve-Compete} paradigm to smartly solve path constraints and a new test case validator to refine the unreliable results produced by LLMs.

\begin{algorithm}[t]
	\caption{{Test Case Validator}} \label{cottontail::alg:validator}
    \small
	\KwIn{a path constraint {\it pc}, a branch {\it br}, a test input from LLM {\it input}, the coverage tree {\it g\_tree}}
	\KwOut{original test input {\it intput} or refined test input {\it input'}, updated global tree {\it g\_tree'}}
	\SetKwFunction{FMain}{\texttt{TestCaseVal}}
    \SetKwProg{Fn}{Function}{:}{}
	\Fn{\FMain{pc, br, input, g\_tree}}{
        res\_eva = {\tt evalauteConstraint}(pc, input) \\
	\If {res\_eva == True}{
            {\tt updateGlobalTree}(g\_tree, br) \\
		\Return input, g\_tree'
	} \Else{
            solution = {\tt getSolution} (pc)\\
            input' = {\tt refineTestCase} (solution, input) \\ 
            {\tt updateGlobalTree}(g\_tree, br) \\
            \Return input', g\_tree'
        }
	}
\end{algorithm}

\subsubsection{LLM-driven Constraint Solver} \label{cottontail::sec:approach:constrint-solver}

It is important to give a precise and logical prompt if we intend to receive output feedback from LLMs (we show that normal prompts generate worse results in \S\ref{cottontail::eva:rq2.2-solver}).
%Therefore, to have a better performance, 
We design a {\it Solve-Complete} paradigm that utilizes the CoT prompt mechanism.
Using CoT prompts instead of normal prompts is advantageous for tasks requiring complex reasoning or multistep problem-solving. 
CoT prompts guide the model to think step-by-step, improving accuracy by reducing errors that arise from skipping intermediate steps. 
It also enhances transparency by explicitly laying out the reasoning process, making it easier to verify the logic and correctness of the solution. 
In summary, the systematic reasoning makes CoT prompts ideal for tackling constraint-solving tasks.

The earlier Figure \ref{cottontail::fig:cot-solving} illustrates a smart constraint-solving strategy grounded in a {\it Solve-Complete} paradigm, where the LLM is asked first to satisfy path constraints and then complete the output to preserve syntactic correctness. 
This process is decomposed into two stages: (1) resolving the {\it Constraint Mask} (`[k!n]')  (e.g., `k!95') using a syntax-aware way by synthesizing a character `e' that satisfies the path constraint under ASCII semantics, and (2) completing the surrounding code such that the entire string remains valid JavaScript to fill the {\it Flexible Mask} (`[xxx]') with a flexible size.
This dual-stage approach mirrors classical symbolic execution techniques, but is uniquely enhanced by the LLM's ability to generate structurally and contextually coherent test inputs. 
In contrast to traditional program synthesis pipelines, which often treat constraint solving and code completion as decoupled steps, this strategy tightly integrates reasoning with generative synthesis. 
The mechanism also aligns with tasks like code infilling, notably benchmarked in CodeXGLUE \cite{lu2021codexglue}, where models are expected to fill in masked code spans while preserving functional correctness. 
However, unlike pure statistical infilling, our approach exhibits explicit constraint awareness, solving constraints before code generation, highlighting the potential of LLMs to unify symbolic reasoning with syntax-preserving code completion. 
In summary, the systematic reasoning capability of the {\it Solve-Compete} paradigm unlocks new applications in symbolic execution for structured test input generation.

%\vspace{1em}

\subsubsection{Test Case Validator} \label{cottontail::sec:approach:testcase-validator}

There is a known issue that LLMs can not reliably generate expected output and can have hallucinations \cite{shankar2024validates,llmSurvey}.
Random output might be acceptable for black/grey box fuzzing, as they do not require the robust (i.e., new test inputs will cover new code coverage) results during each iteration.
However, for a concolic execution, robustness is one of the essential features that should be guaranteed.
Therefore, we need to handle unreliable results produced from GPT to ensure it follows the soundness guarantee of traditional concolic execution and updates the global ECT precisely.

Algorithm \ref{cottontail::alg:validator} presents the workflow of the test case validator, which validates or refines test inputs generated by LLMs. 
The algorithm takes as input a path constraint {\it pc}, a branch {\it br}, the test input ({\it input}) produced by LLMs, and the global coverage tree ({\it g\_tree}). 
It first evaluates whether the input satisfies the path constraint using the function {\tt evaluateConstraint} (Line 2). 
If the constraint is satisfied, i.e., the returned boolean flag {\it res\_eva} is {\it True}, the branch {\it br} is updated in the global tree using {\tt updateGlobalTree} (Line 9), and the algorithm outputs the original test input ({\it input}) alongside the updated tree. 
If the constraint is not satisfied, a solution for the path constraint is computed using {\tt getSolution} (Line 7), and a refined test input ({\it input'}) is generated through function {\tt refineTestCase} (Line 8). Finally, the updated tree ({\it g\_tree'}) is returned along with the refined input. 
In summary, this algorithm ensures the validity of test cases and updates the global coverage to improve test coverage.

In particular, in {\tt refineTestCase} function, we replace the unreliable solution generated by LLM with the correct solution generated by a traditional solver.
By such, even though LLMs produce unreliable outputs, \ourSol could fix them and refine them to the same output as the ones generated by traditional solvers.

\subsection{History-guided Seed Acquisition} \label{cottontail::sec:approach:seed-generation}

In this subsection, we detail the strategy to generate initial seeds before testing or alleviate the saturation issue during testing, including the history coverage recorder and the history-driven seed generator.
Since the key contributions lie in the generation during testing, we detail the history coverage recorder first in the following.

\subsubsection{History Coverage
Recorder} \label{cottontail::sec:approach:coverage-recorder}

It is important to trace the testing history to know {\it which branches} can be covered by {\it what test inputs} and {\it which are the uncovered branches} remaining uncovered. 
By investigating the connection between test input and its covered branches, we could not only understand the underlying processing logic of test programs but also highlight what missing features are within the test inputs.
To practically collect the history information, we continue to leverage the benefits of the informative coverage map (i.e., ECT) to get the covered or uncovered branches. 

After collecting history coverage mappings and extracting branch information from the global coverage map, we then use this information to construct CoT prompts for {\it fresh} seed generation during testing. 

\subsubsection{LLM-driven Seed Generator} \label{cottontail::sec:approach:seed-generator}
The generator is invoked based on two different timings: initial seed acquisition before testing and fresh seed acquisition during testing.

\begin{figure}[t]
  \centering
  \includegraphics[width=0.99\linewidth]{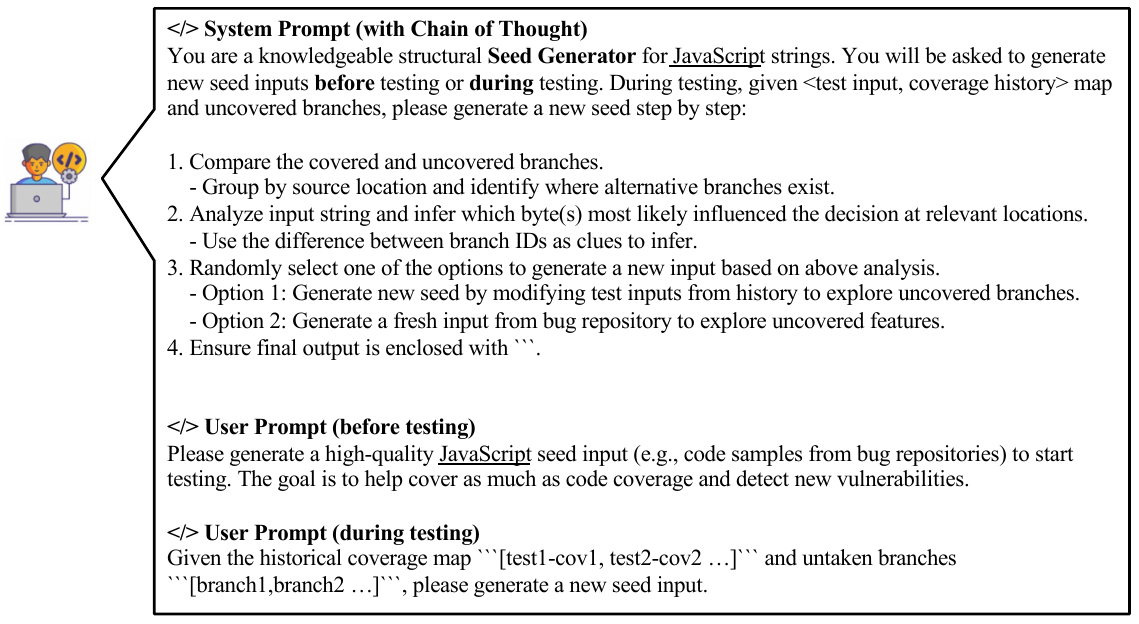}
  %\vspace{-2em}
  \caption{Chain of Thought (CoT) prompts for LLM-driven seed generation %(Sample responses can be found in Figure \ref{cottontail::fig:prompt-seed-generation-response1} and \ref{cottontail::fig:prompt-seed-generation-response2} in Appendix)
  }
  \label{cottontail::fig:prompt-seed-generation}
  %\vspace{-1em}
\end{figure}

\smallskip
\noindent
{\it Initial Seed Acquisition.} 
If there are no interesting seed inputs to set up testing, we define a prompt that helps generate high-quality structured seed inputs.
Since LLMs were trained via tons of code and resources and inspired by many existing studies \cite{xia2024fuzz4all,titanFuzz}, it is reasonable that LLMs have expert knowledge of what kinds of code have triggered vulnerabilities in bug repositories.
Thus, we directly prompt LLMs to generate high-quality structured test inputs from existing bug repositories.
By such, no manual work will be required to collect historical buggy code examples.

\smallskip
\noindent
{\it Fresh Seed Acquisition.} 
During testing, if there is no interesting coverage increase (i.e., saturated) after a timeout (i.e., three minutes), by check the coverage (collected from external tool {\tt gcov}) at runtime.
It is straightforward to apply the same strategy used in seed generation to get a fresh seed, but it is ineffective (demonstrated in \S\ref{cottontail::eva:rq2.4-seed-generation}).
To make it more effective, we design a creative generation solution using CoT (Chain of Thought) prompts to effectively explore the unexplored branches/features during testing. 
Such a design is inspired by an interesting behavior investigated by prior studies, i.e., a better name can help LLMs better understand the program semantics \cite{wang2024doesnamingaffectllms,LLMCodeAnalysis-usenix24}.

Figure \ref{cottontail::fig:prompt-seed-generation} illustrates the prompts designed to guide a seed generator for {\tt JavaScript} in creative generation of seed inputs for different timings.
In particular, the CoT workflow guides an LLM to generate new {\tt JavaScript} seed input through a structured multi-step reasoning process during testing. 
First, the LLM compares covered and uncovered branches, groups them by source location, and identifies divergent execution points.
Then, it analyzes input bytes likely responsible for branching decisions using its internal inferring capacities.
Finally, it synthesizes new inputs either by mutating existing test cases to explore specific branches or by drawing from existing bug repositories.

\begin{figure}[t]
  \centering
  \includegraphics[width=0.99\linewidth]{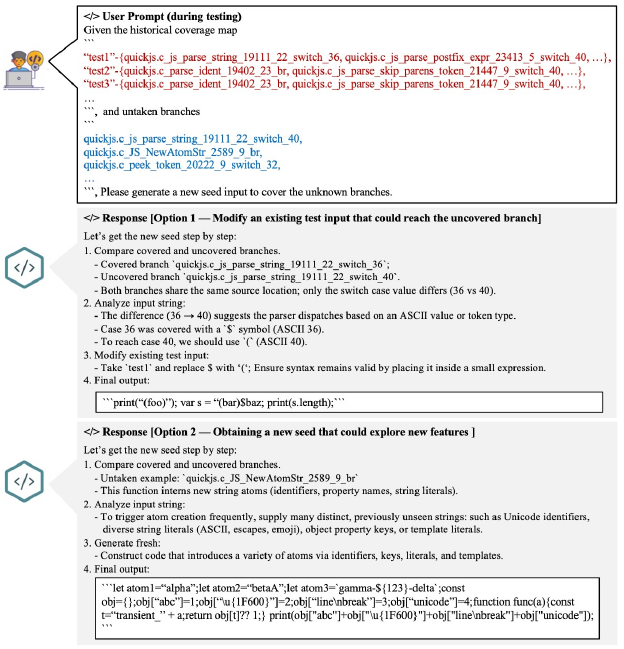}
  \vspace{-1em}
  \caption{Examples of LLM responses of history-guided seed generation}
  \label{cottontail::fig:seed-generation-example}
  \vspace{-1em}
\end{figure}

\smallskip
\noindent
{\bf Examples of LLM-driven Seed Generation.}
To have a better understanding of how history-guided seed acquisition works, we provide two illustrative examples in Figure \ref{cottontail::fig:seed-generation-example} to articulate the seed generation process (for simplicity, only a reduced version is presented).
In principle, \ourSol learns from historical coverage and generates new seeds in two ways: 1) mutating historical test inputs; 2) generating new seeds (or test inputs) from scratch.

In the first response shown in Figure \ref{cottontail::fig:seed-generation-example}, \ourSol generates a new test input by mutating a historical test input that has been executed before. The historical test input is shown in the prompt, and \ourSol asks the LLM to generate a new test input by making some changes to it. The generated test input is similar to the original but contains differences that may lead to the exploration of new program paths.
This strategy is effective because mutating existing test inputs leverages prior knowledge to produce new inputs that are both likely to be valid and potentially interesting.

In the second response, \ourSol generates a new test input from scratch by leveraging the naming conventions used in the target program’s implementation.
This choice is motivated by the expectation that good programmers typically employ meaningful and consistent naming patterns for the intentions of functions \cite{name-methods,how-devs-choose-names, charitsis2022function}. Moreover, recent studies show that LLMs can utilize such identifier information to assist code analysis \cite{LLMCodeAnalysis-usenix24,llm-based-method-name}.

Note that the generated seeds may not always satisfy the requirements for covering the untaken branches. Instead, we expect them to provide useful hints that guide the exploration of the input space, thereby increasing the likelihood of generating a test input that can trigger those branches. If a new seed fails to cover the target branches, \ourSol will quickly discard it guided by the ECT.

\section{Implementation}  \label{cottontail::sec:implementation}

We implemented \ourSol on top of \symcc (commit version 65a3633).
The newly designed components structural instrumentation and coverage-guided path constraints selection (\S\ref{cottontail::sec:approach:constraint-selection}) are implemented as separated functions using C++. The remaining LLM-driven constraint solving (\S\ref{cottontail::sec:approach:constraint-solving}) and history-guided seed acquisition (\S\ref{cottontail::sec:approach:seed-generation}) are implemented in Python code.
The running script is set up using Python 3.9. 
For the setting of different parameters $\alpha$, $\beta$, and $\gamma$ used in Equation \ref{cottontail::eq:selector}, we run extra experiments 
%(presented in Table \ref{cottontail::tab:experiment_results} in the Appendix) 
and opt for one setting that has a better trade-off between efficiency and effectiveness, which is 1.0, 3.0, and 0.8, respectively.
We used {\tt gpt-4o} (only in Setting 1 in RQ1) and {\tt gpt-4o-mini} as our base LLMs and invoked their Python APIs to communicate with the model.
The temperature of the model is set to 0 for better reproducibility.
Note that an LLM with better reasoning capabilities (or higher cost) is preferred but not required.
Our extra experiments show that other recently released cost-effective models (i.e., {\tt deepseek-v3} and {\tt gpt-4.1-nano}) can work very well compared with higher cost models such as {\tt gpt-4o}.
\section{Evaluation}  \label{cottontail::sec:evaluation}

\begin{table}[t]
	\centering
    \small
	\vspace{-1mm}
	\caption{Open source libraries cross four different formats used in evaluation (LOC: lines of code; Stars: GitHub stars)} \label{cottontail::tab:benchmark}
    %\vspace{-1em}
	%\begin{tabular*}{\hsize}{@{}@{\extracolsep{\fill}}ccccc@{}}
	\begin{tabular}{ccccc}
		\toprule
		\multirow{1}{*}{\textbf{Libraries}} &
        \multirow{1}*{\textbf{Format}} & 
		\multirow{1}*{\textbf{Version}} & 
		\multicolumn{1}{c}{\textbf{LOC}} & 
 
		\multicolumn{1}{c}{\textbf{Stars}} \\
		%& &  \textit{(w/o bitmap)} & \textit{(w/ bitmap)} & {\it (w/ CFG)} & {\it (w/ MC)} & {\it (w/ CTSG)} & {\it (w/ ECT)}\\
		%\cmidrule(r){3-14}
		\midrule
		\multirow{1}{*}{\href{https://github.com/GNOME/libxml2}{\textcolor{darkgray}{Libxml2}}} & XML & 2.13.5& 80.0k & 0.6k \\
		\multirow{1}{*}{\href{https://github.com/libexpat/libexpat}{\textcolor{darkgray}{Libexpat}}} & XML &2.6.4& 14.6k& 1.1k \\
		%\multirow{1}{*}{\href{https://github.com/curl/curl}{\textcolor{darkgray}{Curl}}} & URI & 8.11.0 & 41.2k& 36.4k  \\
		%\multirow{1}{*}{\href{https://github.com/lexbor/lexbor}{\textcolor{darkgray}{Lexbor}}} & URI  &2.4.0 & 26.3k & 1.7k \\
		\multirow{1}{*}{\href{https://github.com/sqlite/sqlite}{\textcolor{darkgray}{SQLite }}} & SQL  &3.47.0& 81.3k & 7.0k  \\
		\multirow{1}{*}{\href{https://github.com/symisc/unqlite}{\textcolor{darkgray}{UnQLite}}} & SQL &1.1.9& 22.5k& 2.1k  \\
		\multirow{1}{*}{\href{https://github.com/ccxvii/mujs}{\textcolor{darkgray}{MuJS}}} &  JavaScript &1.3.5& 10.0k & 0.8k  \\
		\multirow{1}{*}{\href{https://github.com/quickjs-ng/quickjs}{\textcolor{darkgray}{QuickJS}}} & JavaScript & 0.7.0 & 46.4k & 1.2k \\
            \multirow{1}{*}{\href{https://github.com/json-c/json-c}{\textcolor{darkgray}{JSON-C}}} & JSON & 0.18& 4.7k  & 3.0k   \\
		\multirow{1}{*}{\href{https://github.com/akheron/jansson}{\textcolor{darkgray}{Jansson}}} & JSON & 2.14& 5.8k & 3.1k  \\
		\bottomrule
	\end{tabular}
	%\vspace{-1em}
\end{table}

\begin{table*}[t]
	\centering
	\caption{Line and branch coverage comparison results against existing concolic execution engines \symcc \cite{SymCC-usenix20} and \marco \cite{hu2024marco}} 
	%\vspace{-1em}
	\begin{tabular}{c} 
	    \includegraphics[width=0.98\linewidth]{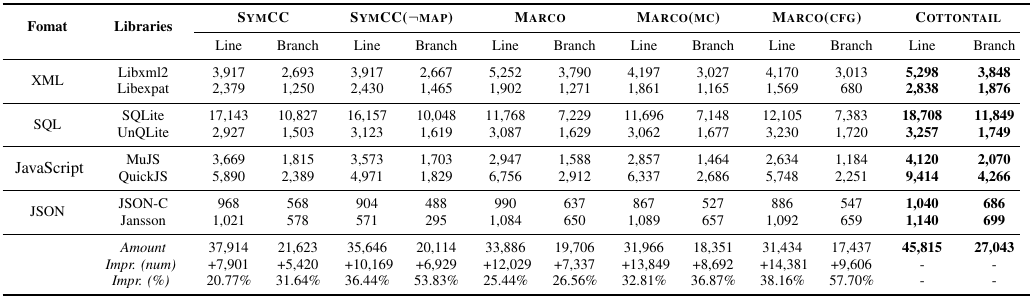}
    \end{tabular}
   \label{cottontail::tab:evaluation:rq1}
    \vspace{-1em}
\end{table*}

To evaluate the effectiveness of \ourSol, we aim to investigate the following research questions (RQs):

\begin{itemize}[leftmargin=1em,nosep]
	\item {\bf RQ1}: How does \ourSol perform compared with state-of-the-art concolic execution approaches?
	%\item {\bf RQ2}: How does \ourSol perform compared with other existing structure-aware fuzzing approaches?
    \item {\bf RQ2}: Can each component contribute to \ourSol?
	\item {\bf RQ3}: Can \ourSol find new vulnerabilities?
        %\item \todo{Add new RQ: {\bf RQ4}: How does \ourSol perform when integrating with open source LLMs?}
\end{itemize}

Among these RQs, RQ1 focuses on demonstrating the effectiveness of \ourSol compared with state-of-the-art approaches and investigating whether \ourSol is superior to them.
RQ2 conducts comprehensive ablation studies to analyze the significance of individual components or key features within \ourSol.
RQ3 assesses the practical vulnerability detection capability of \ourSol. 

All experiments were run on a Linux PC with Intel(R) Xeon(R) W-2133 CPU @ 3.60GHz x 12 processors and 64GB RAM running Ubuntu 18.04 operating system.

%\subsection{Experimental Setup}
\smallskip
\noindent
{\it Benchmarks.} Table \ref{cottontail::tab:benchmark} presents eight widely tested open-source libraries used for evaluation, including libraries for {\tt XML} (Libxml2/Libexpat), {\tt SQL} (SQLite/UnQLite), {\tt JavaScript} (MuJS/QuickJS), and {\tt JSON} (JSON-C/Jansson), varying in size and popularity. 
This diverse set of libraries covers a broad range of formats, codebases, and community adoption levels, making it a comprehensive benchmark suite for evaluation.

\subsection{RQ1: Comparison with Baseline Approaches}
\label{cottontail:eva:rq1}

{\it Comparative Approaches.}
The following state-of-the-art concolic execution approaches are compared:
\begin{itemize}[leftmargin=1em,nosep]
	\item \symcc \cite{SymCC-usenix20}: the tool \ourSol built on top of (enable the {\tt Bitmap}-guided path constraints selection by default).
	\item \symccNoMap: a variant approach of \symcc that selects all newly generated path constraints.
	\item \marco \cite{hu2024marco}: a recent concolic execution engine that constructs CTSG to select path constraints.
	\item \marcoMC: A variant of \marco that adopts Markov Chain modeling in CSTG.
	\item \marcoCFG: A variant of \marco that applies the CFG-directed searching algorithm in CSTG.
\end{itemize}

\begin{figure}[t]
	\centering
	\includegraphics[width=0.99\linewidth]{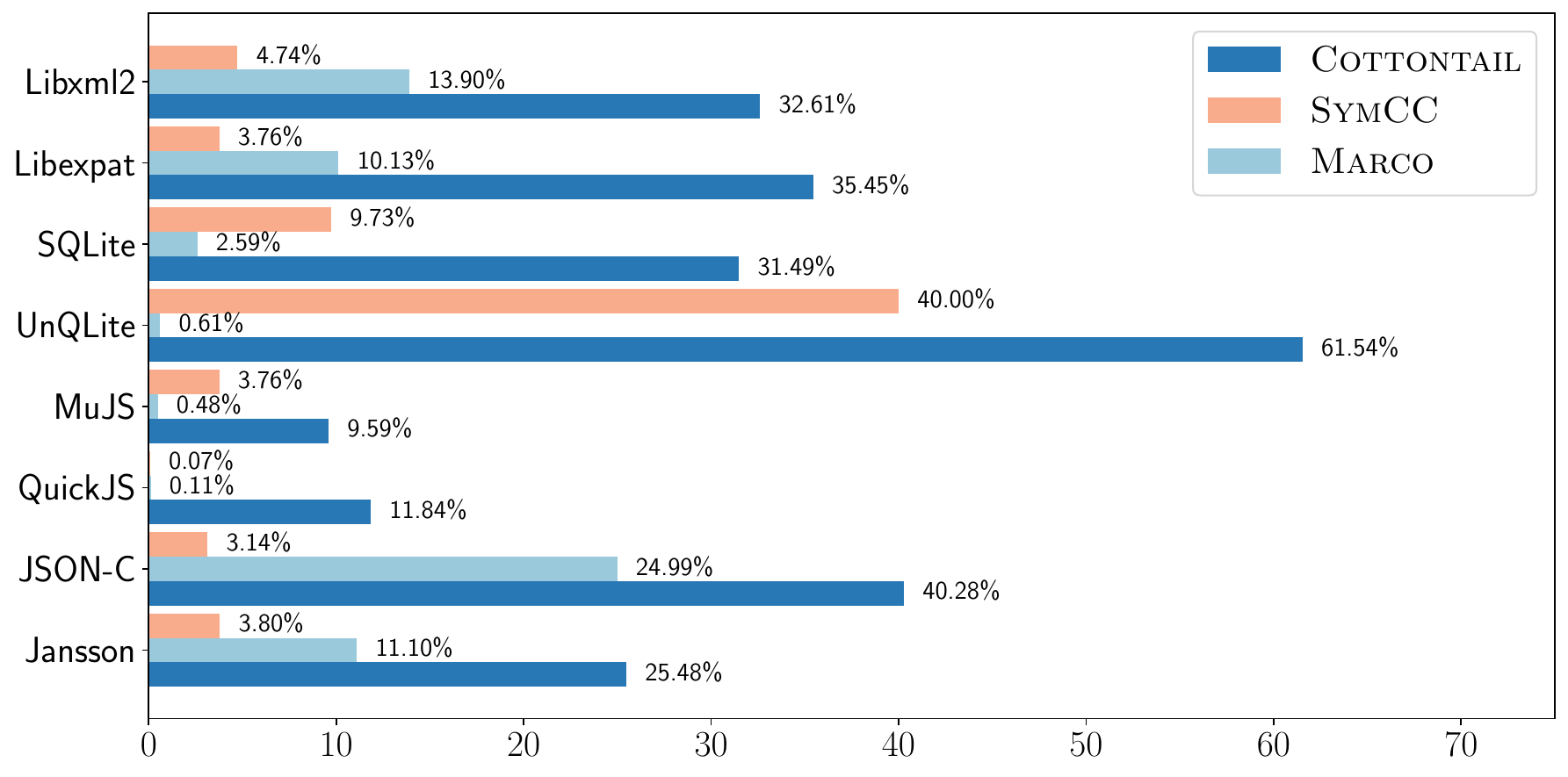}
	%\vspace{-1em}
	\caption{Comparison results of parser checking passing rate (\% in {\it y-axis}) against \symcc and \marco}
	\label{cottontail::fig:evaluation:rq1-passing-rate}
	%\vspace{-1em}
\end{figure}
%* Marco (default 2: full-fledged)
%* Marco-cfg (5)
%* Marco-mc (3)

We select \symcc and \symccNoMap as we built \ourSol on \symcc. 
\marco is an approach proposed recently, and its experiments show that the two variant approaches (i.e., \marcoCFG and \marcoMC) could outperform \marco in a few cases, so we also include them.

\smallskip
\noindent
{\it Running Settings.}
We design three distinct settings to comprehensively demonstrate the superiority of \ourSol.
\begin{itemize}[leftmargin=1em,itemsep=0.1cm]
    \item Setting 1: We run each approach with a timeout of 1 hour.
    \item Setting 2: We run \ourSol with a timeout of 1 hour and other approaches with a timeout of 12 hours.
    \item Setting 3: We set a 12-hour timeout for each approach.
\end{itemize}

We selected a timeout of 1 hour in Settings 1 and 2 for two reasons. 
First, we found that \ourSol not only significantly improves the baseline approaches within one hour, but {\it the results of running our approach for only 1 hour can even beat the baselines running for 12 hours on average}.
Second, as suggested by prior studies \cite{gao2023beyond,klooster2023continuous} and our experimental results in Setting 3, while having longer testing campaigns can boost the effectiveness, we found that the gain when increasing the time budget after 1 hour is overall relatively poor.
We run with a longer running time in Setting 3, as this could help us understand when the testing gets saturated and justify the need for a new seed generation strategy.
\ourSol outperforms baselines in all three settings on average, but due to the page limit, we only report detailed results of Setting 1 in RQ1 and leave other results in the Appendix.

To further conduct a fair comparison, we use the seeds from \marco and launch each tool with the same seeds for all settings.
To help detect possible program issues, we compile the target program built with {\tt AddressSan} \cite{serebryany2012addresssanitizer} and use it as a test oracle to detect memory issues.
In particular, although \ourSol does not require pre-collected seed inputs to set up, for a fair comparison, we disable the seed acquisition contribution in \ourSol. To clarify, the description of \ourSol in this subsection refers to \ourSolNoSeedGen, the variant version of \ourSol where the same initial seeds as the baselines are used and without new seed generation (please check different versions of \ourSol in \S\ref{cottontail::eva:rq2.4-seed-generation}). To reduce the threats from randomness, we repeated running them five times and reported the median results.

\smallskip
\noindent
{\it Metrics.} We use code coverage, including line and branch coverage measured by the external tool {\tt gcovr} to compare the effectiveness of comparative approaches.

\smallskip
\noindent
{\it Results.}
Table \ref{cottontail::tab:evaluation:rq1} provides a comprehensive comparison results achieved by comparative approaches. 
Notably, \ourSol significantly achieves a superior code coverage on average, from 20.77\% to 38.16\% (30.73\% on average and up to 99.65\% improvement over QuickJS) in terms of line coverage and 25.56\% to 57.70\% (41.32\% on average and up to 175.88\% over Libexpat) in terms of branch coverage against comparative approaches, yielding up to 12k more lines and 9k more branches in total.
Note that the very recent approach \marco can only achieve 13\% code coverage than baseline approaches.
Furthermore, in both Settings 2 and Setting 3, \ourSol consistently outperforms all comparative approaches on average, demonstrating the stronger capabilities on code coverage.

To have a better understanding of why \ourSol is superior, we further analyze the validity of generated test cases among \symcc, \marco, and \ourSol.
Since \symcc and \marco produced millions of test cases in 12 hours, we ran them in another 1-hour setting for the same running time.
Figure \ref{cottontail::fig:evaluation:rq1-passing-rate} presents a detailed comparative analysis of parser checking passing rates for three comparative tools. 
From the figure, we can observe that \ourSol consistently performs better in several critical libraries: it achieves a significant 32.61\% passing rate in Libxml2, yielding 588\% higher rate than \symcc's 4.74\% and 140\% higher rate than \marco's 13.90\%.
The overall results suggest that the increased number of valid test inputs helps yield better code coverage results.

\vspace{0.1cm}
\begin{mdframed}[backgroundcolor=gray!15] 
   \vspace{0.01cm}
   \noindent
   {\bf Answer to RQ1:} \ourSol significantly improves the state-of-the-art approaches in terms of line/branch coverage on average in all three running settings, demonstrating the effectiveness of \ourSol in generating highly structured test inputs.
\end{mdframed}

\subsection{RQ2: Ablation Studies}
\label{cottontail::sec:rq2.1}

This subsection presents the methodologies to evaluate the impact of the newly designed components. %\footnote{Due to the page limit, we omit the evaluation on the overhead of structural instrumentation here and present it in Appendix-Section-3.}.

\begin{table*}[t]
    \centering
    \footnotesize
	%\vspace{-1mm}
	\caption{Results of path constraint selector design in \ourSol}
        %\vspace{-1em}
	\begin{tabular}{ccccccc||cccccc}
	\toprule
        \multirow{2}*{\textbf{Libraries}} & 
		\multicolumn{6}{c||}{\textbf{Comparison of Path Constraints Selection --- First Phase}} & 
        \multicolumn{6}{c}{\textbf{Comparison of Path Constraints Selection --- Second Phase}}  \\
		\cmidrule(r){2-13}
            & \multicolumn{2}{c}{NoMap} & \multicolumn{2}{c}{ Bitmap} & \multicolumn{2}{c||}{ECT (ours)} & 
            \multicolumn{2}{c}{ NoMap} & \multicolumn{2}{c}{Bitmap} & \multicolumn{2}{c}{ECT (ours)} \\
            \cmidrule(r){2-13}
            & {\it No.pc} & {\it Cover.} & {\it No.pc} & {\it Cover.} & {\it No.pc} & {\it Cover.} & {\it No.pc} & {\it Cover.} & {\it No.pc} & {\it Cover.} & {\it No.pc} & {\it Cover.}  \\ 
		%\midrule
            \midrule
		\multirow{1}{*}{Libexpat} & 461&1,441 & 187&1,441 & 159&1,441  & 6,149&1,684 & 615&1,681  & 685 &1,687\\
            \multirow{1}{*}{SQLite} & 401 &11,331 & 191&11,331  & 137&11,331  & 3,618& 11,565 & 313 & 11,584 & 687 &11,568  \\
            \multirow{1}{*}{MuJS} & 486&1,404 & 185&1,404  & 273&1,404  & 7,688&2,758 & 653&2,666   & 2,550&2,743 \\
             \multirow{1}{*}{JSON-C} & 223&566 & 106&566  & 67&566 & 7,024&899& 355&861  & 926&887\\
              %\multirow{1}{*}{Jansson} & 242 & 50 & 33 & 34.00\% & 786 & 786 & 1331 \\
		\bottomrule
	\end{tabular} 
        \begin{tablenotes}
        %\footnotesize
        \item * The number {\it A(B)} in the table represents the number of path constraints ({\it No.pc}) collected in the first iterative run (A) and the line coverage ({\it Cover.}) achieved (B). We omitted the comparison with CSTG as it does not follow the iteration working style.
      \end{tablenotes}
	 	%\vspace{-1.5em}
	\label{cottontail::tab:evaluation:rq2.2-selector}
\end{table*}

\smallskip
\noindent
\textbf{\textit{RQ2.1: How effective is the ECT-guided path constraint selection?}} \label{cottontail::eva:rq2.1-path-collection}
As mentioned in \S\ref{cottontail::sec:approach:constraint-selector}, we design two phases to remove redundant path constraints across single concolic execution and in-between runs.
We here evaluate how many path constraints were filtered out in the two phases (e.g., single or in-between concolic execution), comparing with existing selection strategies (i.e., select all without map {\tt Nomap} and guided by {\tt Bitmap}).

To do so, we run the seed input and terminate it after the first iteration is done to evaluate the effectiveness in the first phase.
Then, we run the seed input within two iterations to evaluate the effectiveness of the in-between runs.
Finally, we count the total number of path constraints and code coverage achieved by different selection strategies in the two phases.

The results in Table \ref{cottontail::tab:evaluation:rq2.2-selector} demonstrate the effectiveness of ECT in guiding path constraint selection across both testing phases.
In the first phase, ECT significantly reduces the number of collected path constraints compared to both {\tt NoMap} and {\tt Bitmap} (e.g., 67 vs. 223 and 106 in JSON-C; 137 vs. 401 and 191 in SQLite), while maintaining identical line coverage, indicating that ECT effectively filters redundant constraints without sacrificing exploration. 
In the second phase, ECT continues to show a substantial reduction in the number of path constraints relative to {\tt NoMap} (e.g., 926 vs. 7,024 in JSON-C; 687 vs. 3,618 in SQLite), yet it retains more path constraints than {\tt Bitmap}, enabling it to achieve better coverage than {\tt Bitmap} and comparable or even slightly improved coverage over {\tt NoMap} in most benchmarks. These results highlight ECT's superiority in balancing structure-aware constraint selection.
The superiority is reasonable as AFL's {\tt Bitmap} tends to miss interesting coverage due to hash collisions, limited granularity, and lack of path sensitivity, all of which cause distinct behaviors to appear identical, reducing fuzzing effectiveness \cite{gan2018collafl,manes2019art}, while using {\tt Nomap} will lead to inefficient testing.
Note that the size of ECT depends on the complexity of the test programs. For example, the size is about 58.3KB for {\tt JSON-C} after 12 hours of running.

\smallskip
\noindent
\textbf{\textit{RQ2.2: How effective are the CoT prompts in {\it Solve-Complete} paradigm?} \label{cottontail::eva:rq2.2-solver}}
We have shown the superior performance of LLM-driven constraint solving in Figure \ref{cottontail::fig:llm-solving-intro}.
To better understand the benefits of the CoT prompt, we compare \ourSol with \ourSolNormPro, a variant of \ourSol that removes the CoT prompt.

\begin{figure}[t]
	\centering
	\includegraphics[width=0.98\linewidth]{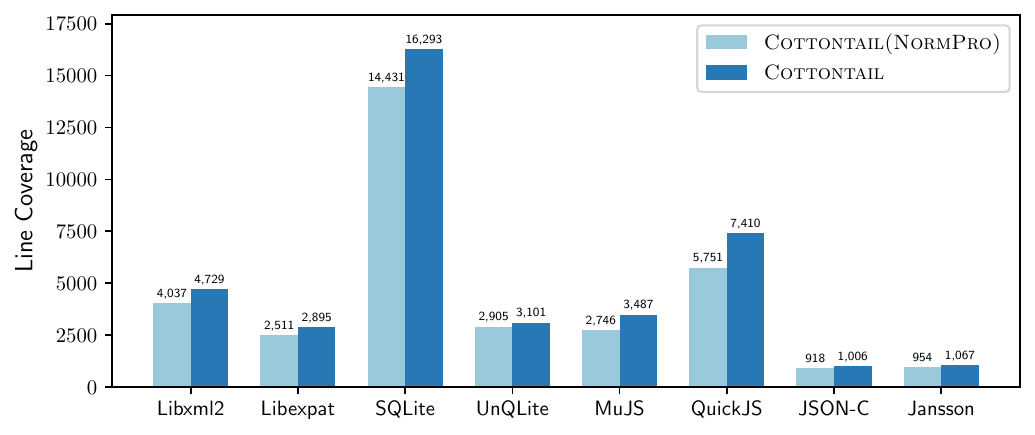} 
	%\vspace{-0.5em}
	\caption{Results of normal and CoT prompts for constraint solving}
        \label{cottontail::fig:evaluation:rq2.3-cot}
	%\vspace{-1em}
\end{figure}

The results in Figure \ref{cottontail::fig:evaluation:rq2.3-cot} highlight the effectiveness of Chain-of-Thought (CoT) prompts in improving constraint solving for line coverage across a diverse set of libraries. 
In all cases, \ourSol with CoT prompts (dark bars) achieves higher coverage than the variant using normal prompts (light bars), with particularly notable improvements observed in SQLite, QuickJS, and MuJS. The substantial gain in SQLite, where coverage increases from approximately 13,000 to over 15,000 lines, underscores how step-by-step reasoning enables the solver to navigate complex constraint spaces better. These results suggest that CoT prompts provide a significant advantage in guiding the model's symbolic reasoning process, leading to more effective exploration and ultimately higher coverage.

\begin{figure}[t]
	\centering
	\includegraphics[width=0.98\linewidth]{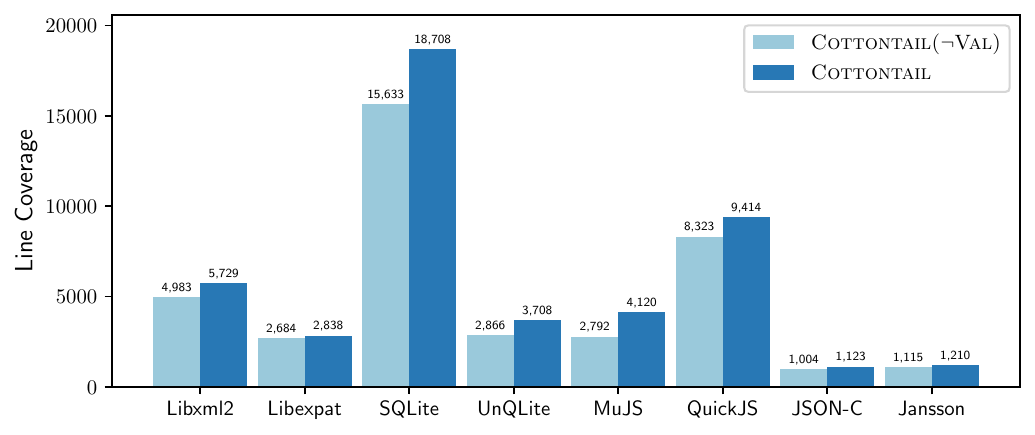}
	\vspace{-0.5em}
	\caption{Comparison results of w/ or w/o test case validator}
	\label{cottontail::fig:evaluation:rq2.3-validator}
	%\vspace{-1em}
\end{figure}

\smallskip
\noindent
\textbf{\textit{RQ2.3: How effective is the test case validator?}} \label{cottontail::eva:rq2.3-validator}
Since it is critical to guarantee the soundness of test cases produced by concolic execution engines, we need to check if the validator designed in \ourSol works.
To have a fair comparison, we measure the line coverage achieved by \ourSol and \ourSolNoValidator (a variant approach of \ourSol that removes the test case validator) under the same setting.

Figure \ref{cottontail::fig:evaluation:rq2.3-validator} presents the line coverage achieved by \ourSol and its variant \ourSolNoValidator across various libraries. 
\ourSol consistently outperforms or matches its counterpart, with significant improvements in larger libraries like SQLite, QuickJS, and Libxml2, where it achieves substantially higher line coverage. 
Overall, the results presented in the figure demonstrate that \ourSol configuration is more effective, especially in complex codebases, highlighting the importance of test case validation and refinement for comprehensive coverage.
This is reasonable as our refinement, designed in Algorithm \ref{cottontail::alg:validator} guarantees the newly generated test cases are expected to explore different program paths. Without the validator, \ourSolNoValidator can be treated as a special variant of smart grey-box fuzzing without the guarantee of systematic program analysis.

We further evaluate the success rate of the LLM-driven constraint solver in directly producing correct solutions. Our results indicate that \ourSol successfully solves 70.08\% of the cases on average, with a failure rate of only 29.92\%. Importantly, this low failure rate does not compromise the soundness of \ourSol, as our newly designed validator—implemented via {\tt refineTestCase} in Algorithm~\ref{cottontail::alg:validator}—systematically refines unreliable results.

\begin{figure*}[t]
	\centering
	\includegraphics[width=1\linewidth]{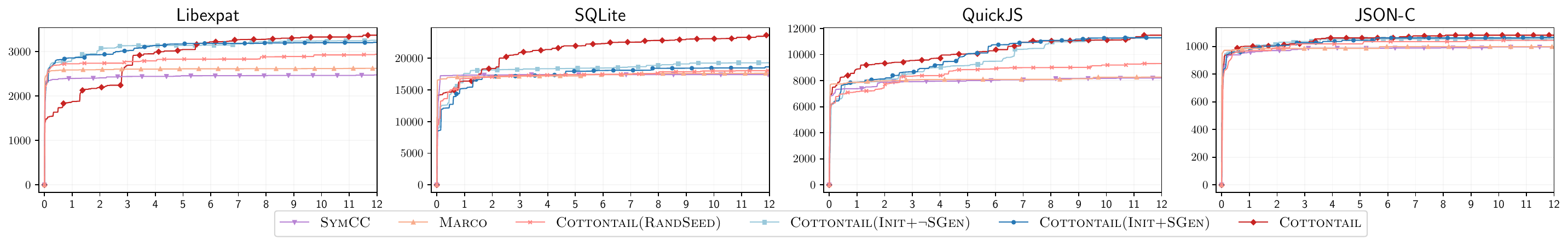}
	\vspace{-1em}
	\caption{Line coverage comparison among \ourSol and variant approaches in 12 hours ({\it x-axis} indicates line coverage while {\it y-axis} the time)}
	\label{cottontail::fig:evaluation:rq2.4}
	%\vspace{-1em}
\end{figure*}

\smallskip
\noindent
\textbf{\textit{RQ2.4: How effective is the history-guided seed acquisition?}} \label{cottontail::eva:rq2.4-seed-generation}
To have a better understanding of the contribution of seed acquisition, we carefully design the following variants:

\begin{itemize}[leftmargin=1em,nosep]
    \item \ourSolRandomSeed: This variant performs random seed generation instead of history-guided generation.
    \item \ourSolNoSeedGen: This variant is run with initial seed inputs and disables seed generation.
    \item \ourSolSeedGen: This variant is run with initial seed inputs and generates new seeds when the test gets saturated (no increased coverage in three minutes).
    \item \ourSol: This is the {\it default} version of our approach, which is run without any initial seed inputs, enabling the history-guided seed acquisition component.
\end{itemize}

We also include the baselines \symcc and \marco to provide a comprehensive comparison. 
We select one benchmark per format and run it for 12 hours to compare its line coverage.
Figure \ref{cottontail::fig:evaluation:rq2.4} presents the detailed results.

\smallskip
\noindent
{\it Contribution of Guided Seed Generation.} 
By comparing the results of \ourSol with \ourSolRandomSeed, we can observe that the history-guided seed acquisition is superior to random seed generation.
In all selected benchmarks, \ourSol consistently outperforms its random-seed variant, achieving noticeably higher line coverage throughout the 12-hour window. For instance, in SQLite, \ourSol reaches over 23,000 lines covered, whereas \ourSolRandomSeed stalls below 19,000.
These results demonstrate that historical execution feedback could guide seed acquisition significantly by prioritizing seeds with higher potential for new coverage.

\smallskip
\noindent
{\it Contribution for Changing Testing Saturation.} 
We conduct two sets of comparative analyses to investigate it.
First, by comparing \ourSolNoSeedGen with \ourSolSeedGen, we can understand how this component boosts testing when the initial seeds are available.
The results show that enabling seed generation significantly improves coverage when an initial seed is available, highlighting the importance of dynamic seed expansion. %\seongmin{The result in Figure 9 seems not supporting this claim: init+nogen looks better than init+gen?}
In particular, we can observe that baseline approaches usually get saturated within 2 hours, and it could be interesting to know how many lines can be covered after the saturation point.
As a result, \ourSolSeedGen covers 101, 376, 806, and 21 more lines over Libexpat, SQLite, QuickJS, and JSON-C than \ourSolNoSeedGen within the latter 10 hours, indicating that enabling the seed generation will continuously increase the coverage, unlike saturating the seed generation.
Second, by comparing \ourSol—including its variant configurations—with \symcc and \marco, we can find out how this component works when there are no seeds. 
The results show that \ourSol maintains superior performance even as coverage begins to saturate, demonstrating its effectiveness in exploring deeper program states under constrained conditions.

\vspace{0.1cm}
\begin{mdframed}[backgroundcolor=gray!15] 
   \vspace{0.01cm}
   \noindent
   {\bf Answer to RQ2:} By conducting carefully designed ablation studies, our results demonstrate the positive contribution of the newly designed components, including structure-aware constraint selection, LLM-driven constraint solving, and history-guided seed acquisition.

\end{mdframed}

\subsection{RQ3: Vulnerability Detection Capability}

\smallskip
\noindent 
{\bf Details of Newly Detected Vulnerabilities.}
To evaluate the practical vulnerability detection capability of \ourSol, we run it (using the setting of the variant approach \ourSolNoSeedGen for a fair comparison) and two baseline approaches in 12 hours and count the number of new vulnerabilities detected.
During the experiments, \ourSol found 6 previously unknown vulnerabilities across three testing subjects and reported them to developers.
The vulnerabilities with their subject, version, short description, and report status are listed in Table \ref{cottontail::tab:evaluation:rq3-vulnerablity}.
These bugs involved heap memory leaks, buffer overflows, and stack overflows, with potential risks such as resource exhaustion, arbitrary code execution, or denial of services. 
Among the detected issues, 4 out of 6 have been fixed when submitting the paper (six new CVE IDs have been assigned), highlighting the practical impact of \ourSol in improving software security.

\smallskip
\noindent
{\it Comparison with Existing Approaches.}
Existing approaches failed or take too much time to detect it due to a structure-agnostic (or heavy) path constraints selection strategy or limited constraint-solving capabilities.
To be specific, \marco can only detect vulnerability \#5. \marco misses the other five vulnerabilities due to the limited path exploration and heavy scheduling on selecting nodes in CSTG. 
For example, when testing MuJS, we found that \marco takes 3.2 out of 12 (26.67\%) hours of computing time to schedule and select an optimal path constraint for solving.
\symcc can only detect four (\#1, \#3, \#5, and \#6) of them and misses others due to the aggressive path constraint elimination and restricted constraint-solving capabilities.

%\ding{56}
%\ding{52}
\begin{table}[t]
    \footnotesize
    %\vspace{-1em}
	\centering
	\caption{New Vulnerabilities Detected} \label{cottontail::tab:evaluation:rq3-vulnerablity}
	%\vspace{-1em}
	\begin{tabular}{cccccc} 
		\toprule
		\textbf{ID}& \textbf{Subject}  & {\bf Description}  & {\bf Status} &\textbf{CVE-Assigned} \\
		\midrule
		\#1 & MuJS & Memory leak  & Fixed & CVE-2024-55061\\
		\#2 & MuJS & Heap overflow  & Fixed & CVE-2025-26082\\
		\#3 & QuickJS  & Stack overflow  & Fixed & CVE-2024-13903\\
		\#4 & QuickJS   & Stack overflow  & Fixed &  CVE-2025-26081\\
		\#5 & UnQLite  & Global overflow  & Reported & CVE-2025-26083\\
		\#6 & UnQLite  & Heap overflow  & Reported & CVE-2025-3791\\
	    \bottomrule
	\label{cottontail::tab:benchmark-cve}
	\end{tabular}
    \vspace{-1.5em}
\end{table}

\smallskip
\noindent 
{\bf Case Study.} \label{cottontail::sec:eva:case-study}
To have a better grasp of the superiority of \ourSol, we present a case study.
%\smallskip
%\noindent
Figure \ref{cottontail::fig:case-study}(a) shows the vulnerable function, and Figure \ref{cottontail::fig:case-study}(b) shows seed and vulnerability triggering input generated by \ourSol.
The issue occurs when {\tt Ap\_sort\_cmp} (Line 2 in Figure \ref{cottontail::fig:case-study}(a)) analyzes the ill-defined comparator (``{\it function(x,y)\{return y;\}}'' shown in Line 4 in Figure \ref{cottontail::fig:case-study}(b)) in the vulnerability triggering input. The unexpected comparator causes {\tt Ap\_sort\_cmp} to access invalid indices, i.e., {\it id\_a} in the array during sorting. After invalid accessing, directly dereferencing the invalid pointer ({\it val\_a}) leads to a heap overflow. In short, the unexpected return value from the {\tt sort} function causes a heap overflow.
Given the seed input\footnote{\scriptsize{\url{https://github.com/unifuzz/seeds/blob/master/general_evaluation/mujs/sort.js}}}, to find a new test input to trigger the overflow, a testing engine should construct an ill-defined {\tt sort} function that returns an unexpected value. The efficient way is to negate the program constraint that requires changing the bytes after the $94{^{th}}$ byte `{\it r}' in {\tt sort} function to a valid {\tt return} statement that returns an unexpected value.

\begin{figure}[t]
\centering
\begin{lrbox}{\mybox}%
\begin{lstlisting}[escapechar=@, frame=single]
// Application logic (buggy function) /* jsarray.c */
static int Ap_sort_cmp(js_State *J, int idx_a,int idx_b){
    js_Object *obj = js_tovalue(J, 0)->u.object;
    if (obj->u.a.simple) {
        js_Value *val_a = &obj->u.a.array[idx_a]; 
        js_Value *val_b = &obj->u.a.array[idx_b];
        int und_a = val_a->t.type == ...; // heap-overflow
    //...
    }
} 
\end{lstlisting}
\end{lrbox}%
\scalebox{0.9}{\usebox{\mybox}}
\vspace{1em}
\centering{ \\ \normalsize {(a) buggy function that triggers a new heap-overflow vulnerability\footnotemark ~detected by \ourSol.}}
\vspace{0.5em}
%// seed input
%c = 30000; a = [];
%for (i = 0; i < 2 * c; i += 1) { a.push(i%c); }
%a.sort(function(x,y){return x - y;}); print(a[2 * c - 2]);

\centering
\begin{lrbox}{\mybox}%
\begin{lstlisting}[escapechar=@]
// LLM generated test input
c = 30000; a = [];
for (i = 0; i < 2 * c; i += 1) {a.push(i%c);} 
a.sort(function (x, y) { r@\colorbox{light-gray}{\makebox(27,3){eturn y;} \}); print(a[100]);}@
\end{lstlisting}
%\vspace{-0.5em}
\end{lrbox}%
\scalebox{0.96}{\usebox{\mybox}}
\vspace{1em}
\centering{\\ \normalsize {(b) Seed input and vulnerability trigger generated by \ourSol (the highlighted \colorbox{light-gray}{strings} are from LLMs).}}
%\vspace{-1em}
\caption{Vulnerable function and triggering input in case study}
\label{cottontail::fig:case-study}
\vspace{-1em}
\end{figure}
\footnotetext{\url{https://github.com/ccxvii/mujs/issues/193}.}

Due to structure-agnostic path constraints selection and limited constraint solving, \marco \cite{hu2024marco} produced 26,575 test inputs (99.9\% invalid) and failed to generate a trigger in 12 hours.
\symcc finds a trigger after 535$^{th}$ iterations while \symccNoMap after 1,811$^{th}$ iterations of constraint solving.
In summary, while existing concolic execution techniques can negate that branch, the resulting input is likely syntactically invalid and requires extra work by the concolic engine to pass the parser checks and generate the syntactically valid input.
%It is worth noting that although they could finally generate a test case to trigger this vulnerability, they may miss important corner cases due to limited computing resources.
In contrast, benefiting from advanced structure-aware path constraint selection and smart constraint solving, \ourSol detects this issue faster within only a few iterations (i.e., 55$^{th}$).

%\vspace{0.1cm}
\begin{mdframed}[backgroundcolor=gray!15] 
   %\vspace{0.01cm}
   \noindent
   {\bf Answer to RQ3:} \ourSol is able to detect previously unknown vulnerabilities, showing a capable practical vulnerability detection capability.
   %\vspace{0.01cm}
\end{mdframed}

%\section{Implications and Discussion}  \label{cottontail::sec:discussion}

%\smallskip
%\noindent 
\section{Perspectives } \label{cottontail::sec:perspectives}

\smallskip
\noindent 
{\bf Potential in Detecting Other Types of Bugs.}
We have shown that the highly structured test inputs could detect previously unknown memory-related vulnerabilities in RQ3. 
The test cases generated by \ourSol could potentially detect other types of bugs (such as parsing or semantic bugs), benefiting from the higher passing rate for parsing checks.
To detect more types of bugs, extra effort may be made to construct well-defined test oracles.
%If a test oracle is well-defined, the test inputs generated by \ourSol could help detect more parser/semantic bugs.
To support our claim, we construct a simple test oracle by differential testing of JSON libraries to detect parsing issues. We define a potential bug as found if two parsers behave differently on the same test input.
Since potential bugs can be false positives, as different parsers may be implemented in different standards (e.g., RFC 4627 for Jansson or RFC 7159 for {\tt JSON-C}), new strategies must be applied to reduce such false positives caused by inconsistent standards.
We manually analyzed a few of the potential issues and found a parsing bug\footnote{\url{https://github.com/json-c/json-c/issues/887}} in JSON-C libraries.
The bug is caused by an incomplete handling of control characters.
Developers have confirmed and fixed the issue we reported.

\smallskip
\noindent 
{\bf Potential in Practical Adoption.}
We believe \ourSol can also have substantial potential to be applied in practical systematic white-box testing, such as SAGE \cite{sage}, from the following four perspectives.
{\it First}, the path constraints that are worth solving are significantly reduced.
As shown in Table \ref{cottontail::tab:evaluation:rq2.2-selector}, the newly designed ECT-based path constraints selection can reduce many redundant path constraints (200\%+ reduction), saving significant testing time in practice.
{\it Second}, the cost of invocation of API is pretty low and \ourSol can be easily integrated with both closed-source and open-source LLMs.
We use the {\tt gpt-4o-mini} as our base LLM, which is an affordable, cheap model (\$0.150 / 1M tokens).
Other LLMs such as {\tt gpt-4.1-nano} (the most cost-effective gpt-4.1 model released on 14/04/2025) and {\tt deepseek-v3} (a cheap and open source model released on 20/01/2025) can also be easily integrated within \ourSol. % (more experimental results are presented in Appendix-Section-4).
{\it Third}, the potential of {\it Solve-Complete}  paradigm for constraint solving is innovative and can be further improved via advanced solutions.
During our experiments, we found that when using the CoT prompts, it would be more beneficial to combine expert knowledge %or more advanced CoT \cite{zhou2022large} 
into the completion phase. 
Our limited knowledge presented in Figure \ref{cottontail::fig:cot-solving} has already shown a significant boost for high-demand structure-aware test input generation in the evaluation. 

\begin{table}[t]
	\centering
    \footnotesize
	%\vspace{-1mm}
	\caption{Time split for execution and constraint solving} \label{cottontail::tab:discussion:solver-time}
    %\vspace{-1em}
	\begin{tabular}{cccccc}
		\toprule
        \multirow{2}*{\textbf{Time Split}} & 
		\multicolumn{2}{c}{SQLite} & 
         \multicolumn{2}{c}{QuickJS} \\
		%& &  \textit{(w/o bitmap)} & \textit{(w/ bitmap)} & {\it (w/ CFG)} & {\it (w/ MC)} & {\it (w/ CTSG)} & {\it (w/ ECT)}\\
		\cmidrule(r){2-5}
		& {\it 4o-mini} & {\it 4.1-nano}   & {\it 4o-mini} & {\it 4.1-nano}  \\
		%\midrule
        \midrule
		\multirow{1}{*}{Execution Time (\%)}  & 9.55 & 12.12 & 10.11 & 11.90\\
		\multirow{1}{*}{Solving Time (\%)} & 90.45 & 87.88  & 89.89 & 88.10\\
		\bottomrule
	\end{tabular} 

	 \vspace{-1em}
	\label{cottontail::tab:rq1}
\end{table}

{\it Fourth}, we investigate the time spent on execution and constraint solving when running \ourSol.
The solving time refers to the time from the engine to take an input to perform concolic execution and give out the solution after constraint solving, and the execution time refers to the time for the execution of the test program with generated test inputs.
Table \ref{cottontail::tab:discussion:solver-time} shows the results over two test programs across two LLMs within 1 hour of concolic testing. 
We could see that the solving time accounts for 89.08\% of the overall testing time.
This is mainly because the GPT API invocation takes time, and it is known that the LLMs are not as fast as traditional solvers like {\tt Z3} \cite{chatafl-ndss24,llmSurvey,cot-llms}. 
Note that such a proportion is reasonable, as constraint solving is a complex process that requires significant computational resources and time \cite{klee,chipounov2011s2e,yun2018qsym-usenix18}. 
For example, both \symcc \cite{SymCC-usenix20} and S2E \cite{chipounov2011s2e} spend more than 90\% of their solving time on analyzing complex software systems (e.g., OpenJPEG). 
Since practical performance is still bound to theoretical limits like constraint solving, 
further improvements (e.g., \cite{accelerate-llm}) on speeding up API invocations could alleviate this issue, as we can already see that a newer version LLM {\tt gpt-4.1-nano} could act faster compared with the older version of LLM.

\section{Discussion}  \label{cottontail::sec:discussion} 

\noindent
{\bf Comparison with Structure-aware Black/Grey-box Fuzzing Approaches.}
In addition to our comparison with state-of-the-art white-box (concolic execution) approaches in the evaluation, we further showcase \ourSol against complementary approaches like structure-aware black/grey-box fuzzing techniques.
%Such a comparison highlights how \ourSol performs relative to them.
To do so, we evaluate \ourSol against an LLM-based black-box fuzzer Fuzz4All \cite{xia2024fuzz4all} and a grey-box fuzzer Nautilus \cite{aschermann2019nautilus} over the same subject {\tt QuickJS}.
We select JavaScript as the target language since it has been employed in the evaluation of Nautilus and is a widely used, complex input format.
Note that Fuzz4All does not support any format that Cottontail supports out of the box. Hence, we implemented additional support for JavaScript in Fuzz4All.
we also used {\tt gpt-4.1} (same model used in \ourSol) and enabled an OpenAI key to set up the autoprompting in Fuzz4All.
For a fair comparison, no initial seed inputs were given for any of the three approaches. 
We use line coverage (collected by {\tt gcov}) to compare the performance of different approaches. 
We repeat fuzzing approaches five times and report median results.

\begin{figure}[t]
	\centering
	%\vspace{-1em}
	\includegraphics[width=1\linewidth]{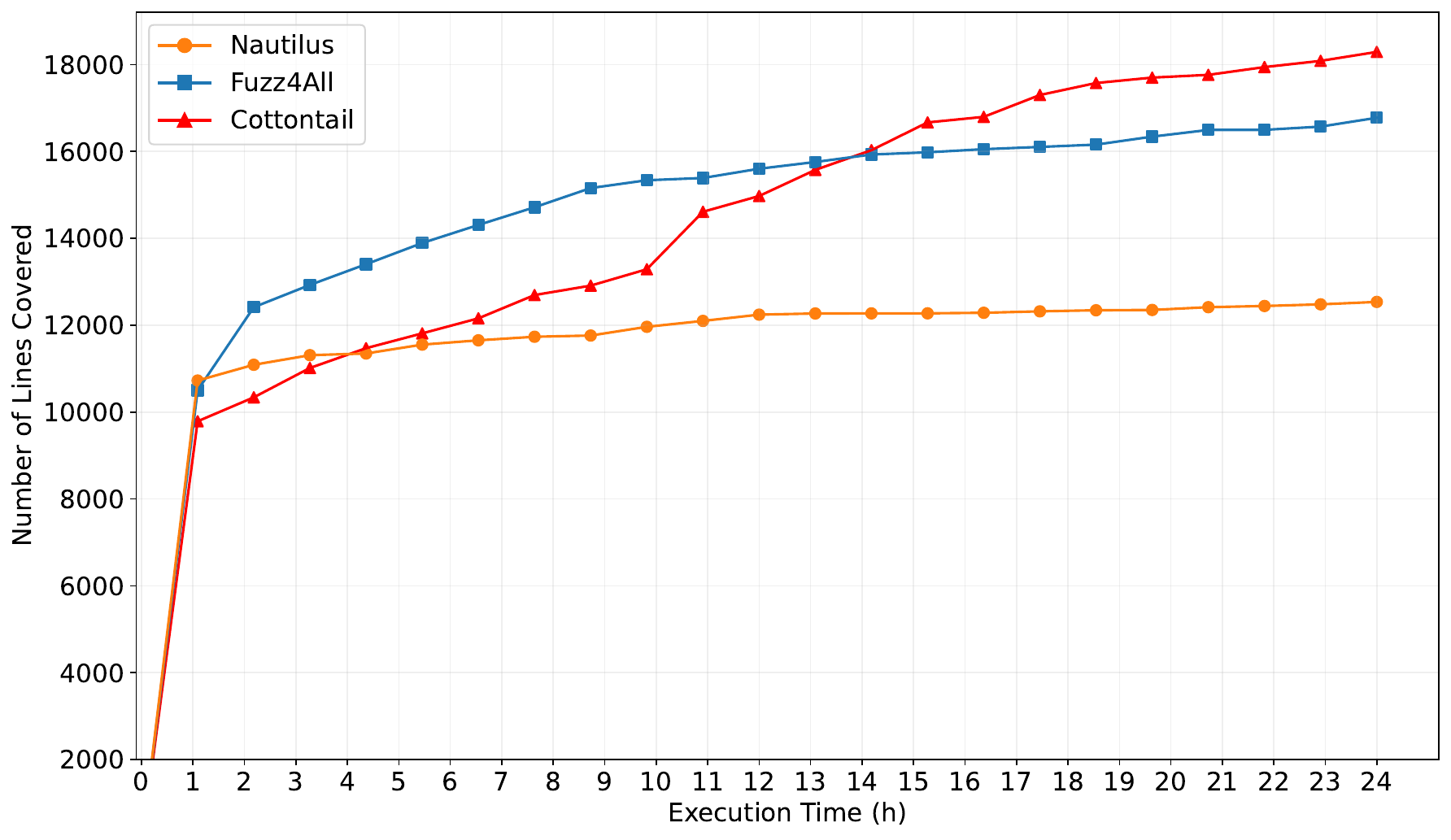} 
	%\vspace{-1em}
	\caption{Comparison results of \ourSol against Fuzz4All \cite{xia2024fuzz4all} and Nautilus \cite{aschermann2019nautilus} over {\tt QuickJS} in 24 hours.}
		\label{cottontail::fig:24h-expr}
	%\vspace{-1em}
\end{figure}

The coverage results are shown in Figure~\ref{cottontail::fig:24h-expr}.
From the figures, we can observe that \ourSol outperforms both Nautilus and Fuzz4All by achieving the highest code coverage when the experiment runs for 24 hours. 
We also notice that \ourSol does not reach the highest coverage at the beginning, as it needs some time to set up concolic execution and gradually generate more test inputs.
However, after a longer run, \ourSol surpasses both baselines, after 4 hours for Nautilus and 14 hours for Fuzz4All.
These results are expected, as although black-box and gray-box fuzzers could act more quickly to generate a large amount of test inputs, they often struggle to effectively explore more of the input space after a longer run (i.e., a saturation point tends to be reached) \cite{klooster2023continuous}.
This is a long-standing challenge for black-box and grey-box fuzzing techniques \cite{fuzzing-challenges}, which tend to get stuck in local optima and fail to explore the input space effectively \cite{gao2023beyond,klooster2023continuous}.
In contrast, \ourSol can not only systematically explore the input space but also leverage the power of LLMs to generate new seed inputs, covering more previously unexplored paths and eventually exploring more paths in the long run.

In summary, if the user has only a few hours of testing budget (e.g., 4 hours or less), black or grey-box fuzzing techniques may be more suitable.
However, if the goal is to achieve high coverage over a longer period (more than 12 hours), \ourSol would be a better choice.

\smallskip
\noindent
{\bf API Costs for Running Experiments.}
The average invocation of GPT at 816 calls per subject, with an average cost of 0.78 USD per hour while using {\tt gpt-4o-mini} model, demonstrating that the cost of using GPT APIs for constraint solving is relatively low. 
Traditional methods of constraint solving are limited by the solving capabilities as the aforementioned in previous sections; they produce a large amount of invalid test cases that have limited contribution to the testing effectiveness for generating highly structured test inputs, although they are faster.
We believe the response time and robustness of LLM could be improved to further facilitate the test input generation capabilities.

\smallskip
\noindent
{\bf Threats to Validity.} 
Our findings and conclusions are subject to several potential threats to validity.
The first concerns external validity, which relates to the generalizability of our results. As the subject of our study, we only selected \symcc and \marco and their variants, the state-of-the-art approaches for concolic execution. 
As objects of our study, we selected eight widely tested open-source libraries covering diverse domains, including XML, SQL, JavaScript, and JSON, which vary in size and popularity. 
While the subjects used in our evaluation are representative of a broad spectrum of real-world applications, we do not claim that the current \ourSol applies to all software programs.
For example, the current version does not include the evaluation over large software systems (e.g., V8 and GCC compilers).
Such a scalability limitation is common in concolic execution, and we plan to integrate advanced techniques (e.g., selective path exploration \cite{chipounov2011s2e}) to alleviate such a limitation.
Another external validity concern driven by the use of LLMs is the risk of data leakage or memorization by LLMs. We believe this is unlikely, as the constraint-solving process in \ourSol is unconventional and unique, making memorization improbable.
The second threat involves internal validity, which refers to the extent to which the evidence supports the causal relationships claimed in our study. LLMs are known to exhibit the hallucination problem, generating outputs that may lack grounding in reality. However, \ourSol addresses this issue by proposing a test case validator to validate and refine the generated test cases. To further reduce the influence of randomness, we also repeated each experiment five times. 

\smallskip
\noindent
{\bf Limitations.}
\ourSol's effectiveness is limited by the completeness of the fuzzing driver. 
It is well-known that writing an effective fuzzing driver can be a challenging and time-consuming process. We plan to leverage the advanced technique \cite{promptfuzz} to mitigate the limitation in the future.
As a source-code-based concolic execution, the current version of \ourSol can only work for the test program whose source code is available.
If only the binary of the target program is available, our approach cannot be directly applied. 
We plan to further transfer the same idea to \textsc{SymQEMU} \cite{poeplau2021symqemu}, a binary concolic execution that shares the same idea of \symcc, to alleviate the limitation.
Another limitation is that it is unclear how \ourSol will perform when handling programs that rely on unstructured inputs or formats unfamiliar to a pre-trained LLM, which we leave as future work to investigate.

\section{Related Work}  \label{cottontail::sec:related-work}

%This section surveys the most related works, including black-box, grey-box, and white-box fuzzing for highly structured test input generation. 

\begin{figure}[t]
	\centering
	\includegraphics[width=0.98\linewidth]{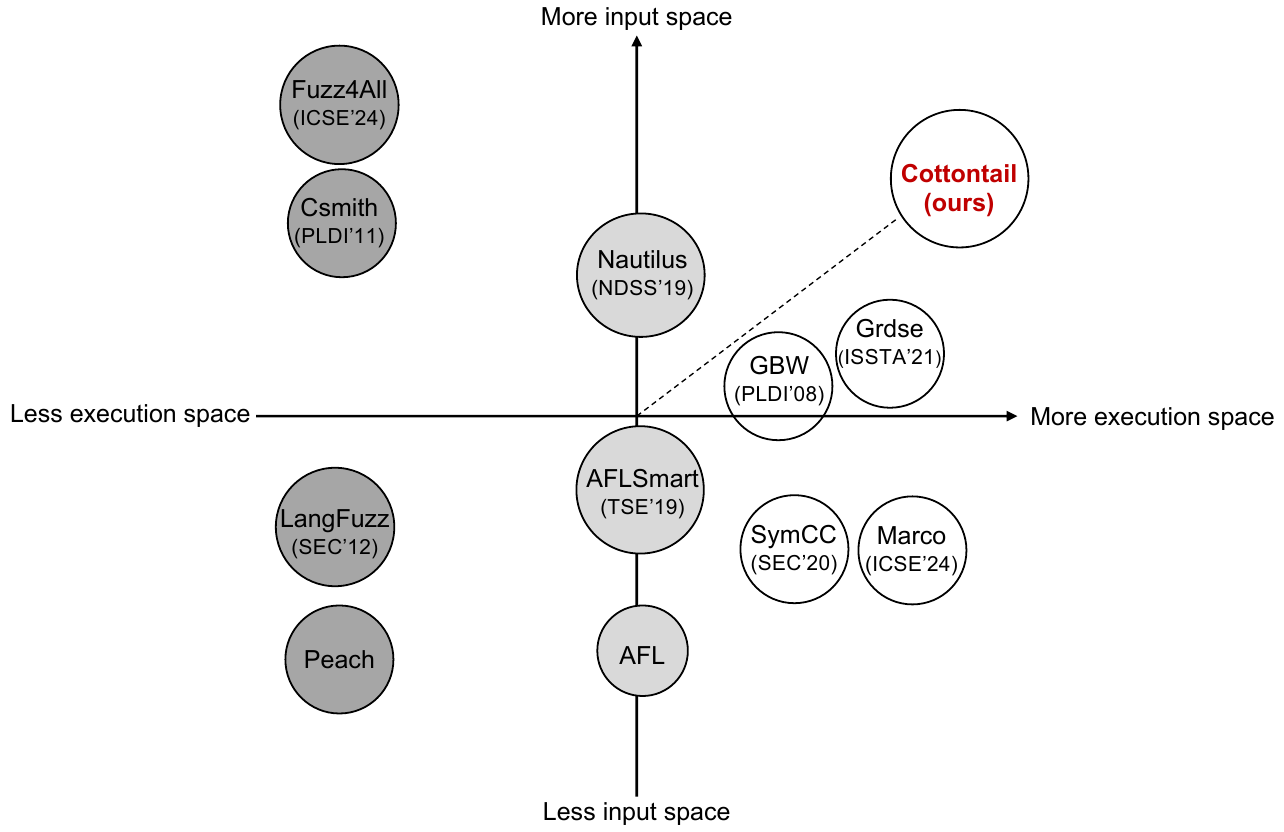}
	%\vspace{-1em}
	\caption{Positioning of \ourSol in exploring input/execution space compared with existing works (the different colors represent three different fuzzing strategies: \tightcolorbox{gray!60}{black-box}, \tightcolorbox{gray!30}{grey-box}, and \tightcolorbox{white!80}{white-box fuzzing}).}
	\label{cottontail::fig:related-work}
	\vspace{-1em}
\end{figure}

\noindent
{\bf Traditional Fuzzing for Software Security.}
In past decades, many fuzzing techniques (including black, grey, and white-box based) have been proposed to improve software security.
Essentially, they aim to explore the input space and execution space of a test program more effectively, where the \emph{input space} refers to the set of all possible inputs that a program can take, and the \emph{execution space} refers to the set of all possible execution paths that a program can run.
To have a clear picture of the positioning of \ourSol, we differentiate our work from existing fuzzing techniques in terms of the capability of exploring input space and execution space in Figure~\ref{cottontail::fig:related-work}.

Black-box fuzzing approaches are typically limited in exploring both input space and execution space, as they do not have access to the internal structure or state of the program being tested.
For example, Peach~\cite{PeachFuzzer} applies format-aware mutations to an initial set of valid inputs using a user-defined input specification.
Black-box grammar-based fuzzers focus on generating inputs that conform to specific syntactic structures, thereby improving the likelihood of covering more input space.
For example, LangFuzz~\cite{holler2012fuzzing} and Grammarinator~\cite{hodovan2018grammarinator} parse existing regression tests using ANTLR grammars for structured input generation.

Greybox fuzzing improves upon black-box techniques by incorporating feedback mechanisms to guide the fuzzing process, which helps in exploring the execution space more effectively.
AFL \cite{fioraldi2020afl++} is a well-known general-purpose fuzzing.
Grey-box grammar-aware fuzzers increase the capabilities of exploring execution space. 
Superion~\cite{wang2019superion} extends LangFuzz with coverage feedback, prioritizing mutated seeds that increase coverage. 
AFLSmart~\cite{pham2019smart} addresses this issue by re-parsing each generated input added to the queue, ensuring structural integrity during fuzzing. 
Weizz~\cite{weizz-ISSTA20} identifies fields and chunks within chunk-based file formats, and NestFuzz~\cite{deng2023nestfuzz} infers inter-field dependencies and the hierarchical structure of inputs for better test case generation.
While grey-box fuzzing techniques have made significant strides in exploring input spaces, they still face challenges in comprehensively exploring execution space.

Concolic execution is known for its capabilities to explore program paths systematically.
QSYM~\cite{yun2018qsym-usenix18} alleviates the strict soundness requirements of conventional concolic executors.
Intriguer~\cite{cho2019intriguer-ccs19} further optimizes QSYM's symbolic execution with field-level knowledge. 
Angora~\cite{angora2018} and Matryoshka~\cite{chen2019matryoshka} opt for taint analysis.
\symcc \cite{SymCC-usenix20} first proposes compilation-based concolic execution to further gain performance enhancement. 
A recent work \marco \cite{hu2024marco} explores code paths by decoupling branch flipping logic from the symbolic tracing logic and defers it until after all branch points uncovered are assessed. 
Syzspec \cite{hao2025syzspec} and Hulk \cite{liu2025domino} are two recent studies that also leverage input specifications to guide concolic execution for better path exploration.
However, traditional concolic execution engines often struggle with generating valid inputs for programs that require highly structured inputs, as they typically do not incorporate knowledge about the input format or syntax.
To address this challenge, grammar-based white-box fuzzing techniques have been proposed.
Godefroid et al. \cite{godefroid2008grammar} introduced a grammar-based white-box fuzzing approach, which advocates for the use of token symbolization during symbolic execution.
The resulting token constraints are then solved using the input grammar. Similarly, CESE \cite{majumdar2007directed} utilizes an input grammar to improve dynamic symbolic execution for programs that parse this grammar. 
Grdse \cite{pan2021grammar} propose grammar-agnostic dynamic symbolic execution that automatically infers input grammars from seed inputs.
Although promising, they are still limited in exploring complex input formats and deep program paths.

%SymFuzz~\cite{cha2015program} identifies dependencies among bit positions in input through symbolic analysis.

Compared with existing fuzzing approaches, we position \ourSol as a novel LLM-driven concolic execution engine that is able to effectively explore both input space and execution space, as shown in Figure~\ref{cottontail::fig:related-work}.
Compared with black/grey-box fuzzing, \ourSol performs systematic path exploration instead of random test case generation/mutation, making it more suitable for comprehensive program analysis.
Compared to white-box fuzzing, \ourSol is superior for its {\it structure-aware path selection}, {\it smart constraint solving}, and {\it capable of acquiring new seeds}, which address three long-standing issues in existing concolic execution approaches. The major contribution, i.e., smart constraint solving that leverages LLM with {\it Solve-Complete} paradigm, is %advanced and 
orthogonal to existing white-box fuzzing techniques, which can be potentially integrated to further improve the performance of concolic execution.

\smallskip
\noindent
{\bf LLM-assisted Fuzzing for Software Security.}
Recent research has demonstrated their potential in software security tasks such as fuzz testing.
ChatFuzz~\cite{chatafl-ndss24} employs an LLM to enhance input generation for protocol fuzzing. 
Codamosa~\cite{lemieux2023codamosa} uses LLMs to generate example test cases for under-covered functions, addressing coverage plateaus in search-based software testing. 
%LLAMAFUZZ~\cite{llamafuzz} extends LLM mutation capabilities to generate valuable structured binary inputs.
Fuzz4All~\cite{xia2024fuzz4all} combines LLMs with evolutionary algorithms to generate structured test inputs for programs C/C++.
CovRL-Fuzz~\cite{eom2024covrl} and InputBlaster~\cite{inputblaster} integrate LLMs to enhance input generation for fuzzing in JavaScript engines and mobile apps.
AutoExe \cite{li2025large} uses a generic code-based representation and performs program synthesis to generate test cases.

Beyond test input generation, LLMs are also applied to generate fuzz drivers for APIs. 
Oss-fuzz-gen~\cite{oss-llm, oss-fuzz-gen} employs few-shot learning techniques to create fuzz drivers based on given APIs. Promptfuzz~\cite{promptfuzz} generates fuzz drivers through various API mutations, and Zhang et al.~\cite{zhang2024effective} evaluate different strategies for LLM-based driver generation.  \textit{TitanFuzz}~\cite{titanFuzz} leverages LLMs to generate both harness programs and arguments for fuzzing deep learning libraries.

Unlike existing LLM-based solutions that randomly generate test inputs, which can explore a larger input space but are limited in exploring execution space (e.g., Fuzz4All \cite{xia2024fuzz4all} as shown in Figure~\ref{cottontail::fig:related-work}),
we propose to combine more precise semantic information (i.e., path constraints) with LLM to improve its test case generation capabilities. Besides, we also utilize a novel {\it Solve-Complete} paradigm for smart constraint solving, yielding promising code coverage and vulnerability detection capabilities.

\section{Conclusion}  \label{cottontail::sec:conclusion}

We presented \ourSol, a new LLM-driven concolic execution engine to generate highly structured test inputs for parsing testing.
\ourSol's novelties lie in the design of structure-aware constraint selection to select path constraints that are worth exploring, LLM-driven constraint solving to smartly produce test cases that not only satisfy the path constraints but also align with syntax rules, and history-guided seed acquisition to generate new seed inputs whenever the engine starts testing or the testing process saturates. 
We compared \ourSol with state-of-the-art concolic execution engines, and the results demonstrate the superior performance of \ourSol in terms of code coverage and vulnerability detection capability.
Our study has shown promising potential in combining traditional program analysis with LLMs, calling for more advanced proposals combining LLMs to improve software security.

%\section{Ethical Considerations}
%\rev{newly added in NDSS2026, not counted in the main text pages}
%This work does not involve human subjects, personal data, or interaction with live systems. All experiments were conducted in controlled environments, and all identified vulnerabilities were responsibly disclosed, and CVE identifiers were assigned to reduce potential misuses. We therefore believe this research does not pose ethical or legal risk.

\ifCLASSOPTIONcompsoc
  \section*{Acknowledgments}
\else
  \section*{Acknowledgment}
\fi

We sincerely appreciate Cristian Cadar for his constructive suggestions in improving the article.
We thank the anonymous reviewers and
our shepherd for their insightful feedback and comments.
We also appreciate the developers of {\tt MuJS}, {\tt QuickJS}, {\tt UnQlite}, and {\tt JSON-C} for their prompt confirmation and fixing of our reported issues.

\bibliographystyle{IEEEtran}
\bibliography{reference}

\begin{thebibliography}{10}
\providecommand{\url}[1]{#1}
\csname url@samestyle\endcsname
\providecommand{\newblock}{\relax}
\providecommand{\bibinfo}[2]{#2}
\providecommand{\BIBentrySTDinterwordspacing}{\spaceskip=0pt\relax}
\providecommand{\BIBentryALTinterwordstretchfactor}{4}
\providecommand{\BIBentryALTinterwordspacing}{\spaceskip=\fontdimen2\font plus
\BIBentryALTinterwordstretchfactor\fontdimen3\font minus
  \fontdimen4\font\relax}
\providecommand{\BIBforeignlanguage}[2]{{%
\expandafter\ifx\csname l@#1\endcsname\relax
\typeout{** WARNING: IEEEtran.bst: No hyphenation pattern has been}%
\typeout{** loaded for the language `#1'. Using the pattern for}%
\typeout{** the default language instead.}%
\else
\language=\csname l@#1\endcsname
\fi
#2}}
\providecommand{\BIBdecl}{\relax}
\BIBdecl

\bibitem{SymCC-usenix20}
S.~Poeplau and A.~Francillon, ``{Symbolic Execution with {SymCC}:
  Don{\textquoteright}t Interpret, Compile!}'' in \emph{Proceedings of the 29th
  USENIX Security Symposium (USENIX Security)}, 2020, pp. 181--198.

\bibitem{SymSan-usenix22}
J.~Chen, W.~Han, M.~Yin, H.~Zeng, C.~Song, B.~Lee, H.~Yin, and I.~Shin,
  ``{{SYMSAN}: Time and Space Efficient Concolic Execution via Dynamic
  Data-flow Analysis},'' in \emph{Proceedings of the 31st USENIX Security
  Symposium (USENIX Security)}, 2022, pp. 2531--2548.

\bibitem{hu2024marco}
J.~Hu, Y.~Duan, and H.~Yin, ``{Marco: A Stochastic Asynchronous Concolic
  Explorer},'' in \emph{Proceedings of the 46th IEEE/ACM International
  Conference on Software Engineering (ICSE)}, 2024, pp. 1--12.

\bibitem{yun2018qsym-usenix18}
I.~Yun, S.~Lee, M.~Xu, Y.~Jang, and T.~Kim, ``{QSYM}: A practical concolic
  execution engine tailored for hybrid fuzzing,'' in \emph{27th USENIX Security
  Symposium (USENIX Security)}, 2018, pp. 745--761.

\bibitem{boosting-se}
H.~Tu, ``{Boosting Symbolic Execution for Heap-Based Vulnerability Detection
  and Exploit Generation},'' in \emph{Proceedings of the 45th International
  Conference on Software Engineering: Companion Proceedings (ICSE-NIER)}, 2023,
  pp. 218--220.

\bibitem{krover}
P.~Pitigalaarachchi, X.~Ding, H.~Qiu, H.~Tu, J.~Hong, and L.~Jiang, ``{KRover:
  A Symbolic Execution Engine for Dynamic Kernel Analysis},'' in
  \emph{Proceedings of the 2023 ACM SIGSAC Conference on Computer and
  Communications Security (CCS)}, 2023, pp. 2009--2023.

\bibitem{fastklee}
H.~Tu, L.~Jiang, X.~Ding, and H.~Jiang, ``{FastKLEE: Faster Symbolic Execution
  via Reducing Redundant Bound Checking of Type-Safe Pointers},'' in
  \emph{Proceedings of the ACM Joint European Software Engineering Conference
  and Symposium on the Foundations of Software Engineering (ESEC/FSE)}, 2022,
  pp. 1741--1745.

\bibitem{lu2021codexglue}
S.~Lu, Z.~Feng, D.~Guo, S.~Wang, D.~Tang, N.~Duan, M.~Zhou \emph{et~al.},
  ``{CodeXGLUE: A Benchmark Dataset and Open Challenge for Code
  Intelligence},'' \emph{arXiv preprint arXiv:2102.04664}, 2021.

\bibitem{nam2024using}
D.~Nam, A.~Macvean, V.~Hellendoorn, B.~Vasilescu, and B.~Myers, ``{Using an LLM
  to Help with Code Understanding},'' in \emph{Proceedings of the IEEE/ACM 46th
  International Conference on Software Engineering (ICSE)}, 2024, pp. 1--13.

\bibitem{sun2024source}
W.~Sun, Y.~Miao, Y.~Li, H.~Zhang, C.~Fang, Y.~Liu, G.~Deng, Y.~Liu, and
  Z.~Chen, ``{Source Code Summarization in the Era of Large Language Models},''
  in \emph{Proceedings of the IEEE/ACM 47th International Conference on
  Software Engineering (ICSE)}, 2025, pp. 1882--1894.

\bibitem{ma2023lms}
W.~Ma, S.~Liu, Z.~Lin, W.~Wang, Q.~Hu, Y.~Liu, C.~Zhang, L.~Nie, L.~Li, and
  Y.~Liu, ``{LMs: Understanding Code Syntax and Semantics for Code Analysis},''
  \emph{arXiv preprint arXiv:2305.12138}, 2023.

\bibitem{LLMCodeAnalysis-usenix24}
C.~Fang, N.~Miao, S.~Srivastav, J.~Liu, R.~Zhang, R.~Fang, Asmita, R.~Tsang,
  N.~Nazari, H.~Wang, and H.~Homayoun, ``{Large Language Models for Code
  Analysis: Do LLMs Really Do Their Job?}'' in \emph{Proceedings of the 33rd
  USENIX Security Symposium (USENIX Security)}, 2024, pp. 829--846.

\bibitem{cot-llms}
J.~Wei, X.~Wang, D.~Schuurmans, M.~Bosma, B.~Ichter, F.~Xia, E.~H. Chi, Q.~V.
  Le, and D.~Zhou, ``{Chain-of-thought Prompting Elicits Reasoning in Large
  Language Models},'' in \emph{Proceedings of the 36th International Conference
  on Neural Information Processing Systems (NeurIPS)}, 2022, pp. 1--14.

\bibitem{CGC}
\BIBentryALTinterwordspacing
D.~of~CGC. {DARPA Cyber Grand Challenge}. [Online]. Available:
  \url{https://www.darpa.mil/research/programs/cyber-grand-challenge}
\BIBentrySTDinterwordspacing

\bibitem{mayhem}
S.~K. Cha, T.~Avgerinos, A.~Rebert, and D.~Brumley, ``{Unleashing Mayhem on
  Binary Code},'' in \emph{Proceedings of the IEEE Symposium on Security and
  Privacy (S\&P)}, 2012, pp. 380--394.

\bibitem{AIxCC}
\BIBentryALTinterwordspacing
D.~of~the Team~Atlanta. {Concolic Execution used in the Winner Team of AIxCC}.
  [Online]. Available:
  \url{https://github.com/Team-Atlanta/aixcc-afc-atlantis/tree/main/example-crs-webservice/crs-multilang/uniafl/src/concolic}
\BIBentrySTDinterwordspacing

\bibitem{z3}
\BIBentryALTinterwordspacing
Z3. (2025) {A Theorem Prover from Microsoft Research}. [Online]. Available:
  \url{https://github.com/z3prover/z3}
\BIBentrySTDinterwordspacing

\bibitem{symsize}
D.~Trabish, S.~Itzhaky, and N.~Rinetzky, ``A bounded symbolic-size model for
  symbolic execution,'' in \emph{Proceedings of ACM Joint Meeting on European
  Software Engineering Conference and Symposium on the Foundations of Software
  Engineering}, 2021, pp. 1190--1201.

\bibitem{SymLoc}
H.~Tu, L.~Jiang, J.~Hong, X.~Ding, and H.~Jiang, ``{Concretely Mapped Symbolic
  Memory Locations for Memory Error Detection},'' \emph{IEEE Transactions on
  Software Engineering}, vol.~50, no.~7, pp. 1747--1767, 2024.

\bibitem{promptfuzz}
Y.~Lyu, Y.~Xie, P.~Chen, and H.~Chen, ``{Prompt Fuzzing for Fuzz Driver
  Generation},'' in \emph{Proceedings of ACM SIGSAC Conference on Computer and
  Communications Security}, 2024, pp. 3793--3807.

\bibitem{FuzzGen}
K.~Ispoglou, D.~Austin, V.~Mohan, and M.~Payer, ``{FuzzGen}: Automatic fuzzer
  generation,'' in \emph{Proceedings of the 29th USENIX Security Symposium
  (USENIX Security)}, 2020, pp. 2271--2287.

\bibitem{UTopia}
B.~Jeong, J.~Jang, H.~Yi, J.~Moon, J.~Kim, I.~Jeon, T.~Kim, W.~Shim, and Y.~H.
  Hwang, ``Utopia: Automatic generation of fuzz driver using unit tests,'' in
  \emph{Proceedings of the IEEE Symposium on Security and Privacy (S\&P)},
  2023, pp. 2676--2692.

\bibitem{lin2025large}
J.~Lin and D.~Mohaisen, ``{From Large to Mammoth: A Comparative Evaluation of
  Large Language Models in Vulnerability Detection},'' in \emph{Proceedings of
  the Network and Distributed System Security Symposium (NDSS)}, 2025, pp.
  1--18.

\bibitem{pearce2023examining}
H.~Pearce, B.~Tan, B.~Ahmad, R.~Karri, and B.~Dolan-Gavitt, ``Examining
  zero-shot vulnerability repair with large language models,'' in
  \emph{Proceedings of the IEEE Symposium on Security and Privacy (S\&P)},
  2023, pp. 2339--2356.

\bibitem{concollmic}
Z.~Luo, H.~Zhao, D.~Wolff, C.~Cadar, and A.~Roychoudhury, ``{Agentic Concolic
  Execution},'' in \emph{Proceedings of the IEEE Symposium on Security and
  Privacy (S\&P)}, 2026, pp. 1--19.

\bibitem{LLMEvaSP24}
S.~Ullah, M.~Han, S.~Pujar, H.~Pearce, A.~Coskun, and G.~Stringhini, ``{LLMs
  Cannot Reliably Identify and Reason About Security Vulnerabilities (Yet?): A
  Comprehensive Evaluation, Framework, and Benchmarks},'' in \emph{Proceedings
  of the IEEE Symposium on Security and Privacy (S\&P)}, 2024, pp. 862--880.

\bibitem{llmSurvey}
Y.~Chang, X.~Wang, J.~Wang, Y.~Wu, L.~Yang, K.~Zhu, H.~Chen, X.~Yi, C.~Wang,
  Y.~Wang, W.~Ye, Y.~Zhang, Y.~Chang, P.~S. Yu, Q.~Yang, and X.~Xie, ``{A
  Survey on Evaluation of Large Language Models},'' \emph{ACM Transactions on
  Intelligent Systems and Technology}, vol.~15, no.~3, 2024.

\bibitem{nong2024chainofthoughtpromptinglargelanguage}
\BIBentryALTinterwordspacing
Y.~Nong, M.~Aldeen, L.~Cheng, H.~Hu, F.~Chen, and H.~Cai, ``Chain-of-thought
  prompting of large language models for discovering and fixing software
  vulnerabilities,'' 2024. [Online]. Available:
  \url{https://arxiv.org/abs/2402.17230}
\BIBentrySTDinterwordspacing

\bibitem{xia2024fuzz4all}
C.~S. Xia, M.~Paltenghi, J.~Le~Tian, M.~Pradel, and L.~Zhang, ``{Fuzz4All:
  Universal Fuzzing with Large Language Models},'' in \emph{Proceedings of the
  IEEE/ACM 46th International Conference on Software Engineering (ICSE)}, 2024,
  pp. 1--13.

\bibitem{titanFuzz}
Y.~Deng, C.~S. Xia, H.~Peng, C.~Yang, and L.~Zhang, ``{Large Language Models
  Are Zero-Shot Fuzzers: Fuzzing Deep-Learning Libraries via Large Language
  Models},'' in \emph{Proceedings of the ACM International Symposium on
  Software Testing and Analysis}, 2023, pp. 423--435.

\bibitem{chatafl-ndss24}
R.~Meng, M.~Mirchev, M.~B{\"o}hme, and A.~Roychoudhury, ``{Large Language Model
  Guided Protocol Fuzzing},'' in \emph{Proceedings of the 31st Annual Network
  and Distributed System Security Symposium (NDSS)}, 2024, pp. 1--15.

\bibitem{klee}
C.~Cadar, D.~Dunbar, and D.~Engler, ``{KLEE: Unassisted and Automatic
  Generation of High-Coverage Tests for Complex Systems Programs},'' in
  \emph{Proceedings of the 8th USENIX Conference on Operating Systems Design
  and Implementation (OSDI)}, 2008, pp. 209--224.

\bibitem{stephens2016driller}
N.~Stephens, J.~Grosen, C.~Salls, A.~Dutcher, R.~Wang, J.~Corbetta,
  Y.~Shoshitaishvili, C.~Kruegel, and G.~Vigna, ``{Driller: Augmenting Fuzzing
  Through Selective Symbolic Execution},'' in \emph{Proceedings of the Network
  and Distributed System Security Symposium (NSDI)}, 2016, pp. 1--16.

\bibitem{shankar2024validates}
S.~Shankar, J.~Zamfirescu-Pereira, B.~Hartmann, A.~Parameswaran, and I.~Arawjo,
  ``{Who Validates the Validators? Aligning LLM-assisted Evaluation of LLM
  Outputs with Human Preferences},'' in \emph{Proceedings of the Annual ACM
  Symposium on User Interface Software and Technology (UIST)}, 2024, pp. 1--14.

\bibitem{wang2024doesnamingaffectllms}
\BIBentryALTinterwordspacing
Z.~Wang, L.~Zhang, C.~Cao, N.~Luo, X.~Luo, and P.~Liu, ``{How Does Naming
  Affect LLMs on Code Analysis Tasks?}'' 2024. [Online]. Available:
  \url{https://arxiv.org/abs/2307.12488}
\BIBentrySTDinterwordspacing

\bibitem{name-methods}
R.~S. Alsuhaibani, C.~D. Newman, M.~J. Decker, M.~L. Collard, and J.~I.
  Maletic, ``On the naming of methods: A survey of professional developers,''
  in \emph{Proceedings of the 43rd International Conference on Software
  Engineering (ICSE)}, 2021, pp. 587--–599.

\bibitem{how-devs-choose-names}
D.~G. Feitelson, A.~Mizrahi, N.~Noy, A.~B. Shabat, O.~Eliyahu, and R.~Sheffer,
  ``{How Developers Choose Names},'' \emph{IEEE Transactions on Software
  Engineering}, vol.~48, no.~01, pp. 37--52, 2022.

\bibitem{charitsis2022function}
C.~Charitsis, C.~Piech, and J.~C. Mitchell, ``Function names: Quantifying the
  relationship between identifiers and their functionality to improve them,''
  in \emph{Proceedings of the Ninth ACM Conference on Learning@ Scale}, 2022,
  pp. 93--101.

\bibitem{llm-based-method-name}
W.~Akram, Y.~Jiang, Y.~Zhang, H.~A. Khan, and H.~Liu, ``{LLM-Based Method Name
  Suggestion with Automatically Generated Context-Rich Prompts},''
  \emph{Proceedings of the ACM Software Engineering}, vol.~2, no. FSE, pp.
  1--22, 2025.

\bibitem{gao2023beyond}
W.~Gao, V.-T. Pham, D.~Liu, O.~Chang, T.~Murray, and B.~I. Rubinstein,
  ``{Beyond the Coverage Plateau: A Comprehensive Study of Fuzz Blockers
  (Registered Report)},'' in \emph{Proceedings of the 2nd International Fuzzing
  Workshop}, 2023, pp. 47--55.

\bibitem{klooster2023continuous}
T.~Klooster, F.~Turkmen, G.~Broenink, R.~Ten~Hove, and M.~B{\"o}hme,
  ``{Continuous Fuzzing: A Study of the Effectiveness and Scalability of
  Fuzzing in CI/CD Pipelines},'' in \emph{IEEE/ACM International Workshop on
  Search-Based and Fuzz Testing (SBFT)}, 2023, pp. 25--32.

\bibitem{serebryany2012addresssanitizer}
K.~Serebryany, D.~Bruening, A.~Potapenko, and D.~Vyukov, ``{AddressSanitizer: A
  Fast Address Sanity Checker},'' in \emph{Proceedings of the USENIX Annual
  Technical Conference}, 2012, pp. 309--318.

\bibitem{gan2018collafl}
S.~Gan, C.~Zhang, X.~Qin, X.~Tu, K.~Li, Z.~Pei, and Z.~Chen, ``{Collafl: Path
  Sensitive Fuzzing},'' in \emph{Proceedings of the IEEE Symposium on Security
  and Privacy (S\&P)}, 2018, pp. 679--696.

\bibitem{manes2019art}
V.~J. Man{\`e}s, H.~Han, C.~Han, S.~K. Cha, M.~Egele, E.~J. Schwartz, and
  M.~Woo, ``{The Art, Science, and Engineering of Fuzzing: A Survey},''
  \emph{IEEE Transactions on Software Engineering}, vol.~47, no.~11, pp.
  2312--2331, 2019.

\bibitem{sage}
P.~Godefroid, M.~Y. Levin, and D.~Molnar, ``{SAGE: Whitebox Fuzzing for
  Security Testing},'' \emph{Communications of the ACM}, vol.~55, no.~3, pp.
  40--44, 2012.

\bibitem{chipounov2011s2e}
V.~Chipounov, V.~Kuznetsov, and G.~Candea, ``{S2E: A Platform for In-vivo
  Multi-path Analysis of Software Systems},'' \emph{ACM SIGPLAN Notices},
  vol.~46, no.~3, pp. 265--278, 2011.

\bibitem{accelerate-llm}
K.~Ayoub. {Accelerating Large Language Models with {TensorRT-LLM} and Serving
  ({OpenAI}-Compatible {API})}.
  \url{https://blog.gopenai.com/accelerating-large-language-models-with-tensorrt-llm-db928323ddbf}.

\bibitem{aschermann2019nautilus}
C.~Aschermann, T.~Frassetto, T.~Holz, P.~Jauernig, A.-R. Sadeghi, and
  D.~Teuchert, ``{NAUTILUS: Fishing for Deep Bugs with Grammars},'' in
  \emph{Proceedings of the Network and Distributed System Security Symposium
  (NDSS)}, 2019, pp. 1--15.

\bibitem{fuzzing-challenges}
M.~Boehme, C.~Cadar, and A.~Roychoudhury, ``{Fuzzing: Challenges and
  Reflections},'' \emph{IEEE Software}, vol.~38, no.~03, pp. 79--86, 2021.

\bibitem{poeplau2021symqemu}
S.~Poeplau and A.~Francillon, ``{{SymQEMU}: Compilation-based Symbolic
  Execution for Binaries},'' in \emph{Proceedings of Network and Distributed
  System Security Symposium (NDSS)}, 2021, pp. 1--18.

\bibitem{PeachFuzzer}
\BIBentryALTinterwordspacing
Peach {{Fuzzer}}. [Online]. Available:
  \url{https://peachtech.gitlab.io/peach-fuzzer-community/}
\BIBentrySTDinterwordspacing

\bibitem{holler2012fuzzing}
C.~Holler, K.~Herzig, and A.~Zeller, ``{Fuzzing with Code Fragments},'' in
  \emph{Proceedings of the 21st USENIX Security Symposium (USENIX Security)},
  2012, pp. 445--458.

\bibitem{hodovan2018grammarinator}
R.~Hodov{\'a}n, {\'A}.~Kiss, and T.~Gyim{\'o}thy, ``{Grammarinator: A
  Grammar-based Open Source Fuzzer},'' in \emph{Proceedings of the 9th ACM
  SIGSOFT International Workshop on Automating TEST Case Design, Selection, and
  Evaluation}, 2018, pp. 45--48.

\bibitem{fioraldi2020afl++}
A.~Fioraldi, D.~Maier, H.~Ei{\ss}feldt, and M.~Heuse, ``{AFL++: Combining
  Incremental Steps of Fuzzing Research},'' in \emph{14th USENIX Workshop on
  Offensive Technologies (WOOT)}, 2020, pp. 1--12.

\bibitem{wang2019superion}
J.~Wang, B.~Chen, L.~Wei, and Y.~Liu, ``{Superion: Grammar-aware Greybox
  Fuzzing},'' in \emph{Proceedings of the IEEE/ACM International Conference on
  Software Engineering (ICSE)}, 2019, pp. 724--735.

\bibitem{pham2019smart}
V.-T. Pham, M.~B{\"o}hme, A.~E. Santosa, A.~R. C{\u{a}}ciulescu, and
  A.~Roychoudhury, ``{Smart Greybox Fuzzing},'' \emph{IEEE Transactions on
  Software Engineering}, vol.~47, no.~9, pp. 1980--1997, 2019.

\bibitem{weizz-ISSTA20}
A.~Fioraldi, D.~C. D'Elia, and E.~Coppa, ``{{WEIZZ}: Automatic Grey-box Fuzzing
  for Structured Binary Formats},'' in \emph{Proceedings of the 29th ACM
  SIGSOFT International Symposium on Software Testing and Analysis (ISSTA)},
  2020, pp. 1--13.

\bibitem{deng2023nestfuzz}
P.~Deng, Z.~Yang, L.~Zhang, G.~Yang, W.~Hong, Y.~Zhang, and M.~Yang,
  ``{NestFuzz: Enhancing Fuzzing with Comprehensive Understanding of Input
  Processing Logic},'' in \emph{Proceedings of the 2023 ACM SIGSAC Conference
  on Computer and Communications Security (CCS)}, 2023, pp. 1272--1286.

\bibitem{cho2019intriguer-ccs19}
M.~Cho, S.~Kim, and T.~Kwon, ``{Intriguer: Field-level Constraint Solving for
  Hybrid Fuzzing},'' in \emph{Proceedings of the ACM Conference on Computer and
  Communications Security}, 2019, pp. 515--530.

\bibitem{angora2018}
P.~Chen and H.~Chen, ``{Angora: Efficient Fuzzing by Principled Search},'' in
  \emph{Proceedings of the IEEE Symposium on Security and Privacy (S\&P)},
  2018, pp. 711--725.

\bibitem{chen2019matryoshka}
P.~Chen, J.~Liu, and H.~Chen, ``{Matryoshka: Fuzzing Deeply Nested Branches},''
  in \emph{Proceedings of the ACM SIGSAC Conference on Computer and
  Communications Security (CCS)}, 2019, pp. 499--513.

\bibitem{hao2025syzspec}
Y.~Hao, J.~Pu, X.~Li, Z.~Qian, and A.~A. Sani, ``{SyzSpec: Specification
  Generation for Linux Kernel Fuzzing via Under-Constrained Symbolic
  Execution},'' in \emph{Proceedings of the ACM SIGSAC Conference on Computer
  and Communications Security (CCS)}, 2025, pp. 1--14.

\bibitem{liu2025domino}
Z.~Liu, T.~Lee, J.~Yu, Z.~Kang, and Y.~Cao, ``{The DOMino Effect: Detecting and
  Exploiting DOM Clobbering Gadgets via Concolic Execution with Symbolic
  DOM},'' in \emph{Proceedings of the 34th USENIX Security Symposium (USENIX
  Security)}, 2025, pp. 8293--8312.

\bibitem{godefroid2008grammar}
P.~Godefroid, A.~Kiezun, and M.~Y. Levin, ``{Grammar-based Whitebox Fuzzing},''
  in \emph{Proceedings of the ACM Conference on Programming Language Design and
  Implementation (PLDI)}, 2008, pp. 206--215.

\bibitem{majumdar2007directed}
R.~Majumdar and R.-G. Xu, ``{Directed Test Generation Using Symbolic
  Grammars},'' in \emph{Proceedings of the 22nd IEEE/ACM International
  Conference on Automated Software Engineering (ICSE)}, 2007, pp. 134--143.

\bibitem{pan2021grammar}
W.~Pan, Z.~Chen, G.~Zhang, Y.~Luo, Y.~Zhang, and J.~Wang, ``{Grammar-Agnostic
  Symbolic Execution by Token Symbolization},'' in \emph{Proceedings of the
  30th ACM SIGSOFT International Symposium on Software Testing and Analysis
  (ISSTA)}, 2021, pp. 374--387.

\bibitem{lemieux2023codamosa}
C.~Lemieux, J.~P. Inala, S.~K. Lahiri, and S.~Sen, ``{Codamosa: Escaping
  Coverage Plateaus in Test Generation with Pre-trained Large Language
  Models},'' in \emph{Proceedings of the IEEE/ACM International Conference on
  Software Engineering (ICSE)}, 2023, pp. 919--931.

\bibitem{eom2024covrl}
J.~Eom, S.~Jeong, and T.~Kwon, ``{Fuzzing JavaScript Interpreters with
  Coverage-Guided Reinforcement Learning for LLM-Based Mutation},'' in
  \emph{Proceedings of the ACM International Symposium on Software Testing and
  Analysis (ISSTA)}, 2024, pp. 1656--1668.

\bibitem{inputblaster}
Z.~Liu, C.~Chen, J.~Wang, M.~Chen, B.~Wu, Z.~Tian, Y.~Huang, J.~Hu, and
  Q.~Wang, ``{Testing the Limits: Unusual Text Inputs Generation for Mobile App
  Crash Detection with Large Language Model},'' in \emph{Proceedings of the
  IEEE/ACM 46th International Conference on Software Engineering (ICSE)}, 2024,
  pp. 1--12.

\bibitem{li2025large}
Y.~Li, R.~Meng, and G.~J. Duck, ``{Large Language Model powered Symbolic
  Execution},'' vol.~9, no. OOPSLA2, 2025, pp. 1--29.

\bibitem{oss-llm}
J.~M. Dongge~Liu and O.~Chang. {Fuzz Target Generation Using LLMs}.
  \url{https://google.github.io/oss-fuzz/research/llms/target_generation/}.

\bibitem{oss-fuzz-gen}
Google, ``oss-fuzz-gen,'' \url{https://github.com/google/oss-fuzz-gen}.

\bibitem{zhang2024effective}
C.~Zhang, Y.~Zheng, M.~Bai, Y.~Li, W.~Ma, X.~Xie, Y.~Li, L.~Sun, and Y.~Liu,
  ``{How Effective Are They? Exploring Large Language Model Based Fuzz Driver
  Generation},'' in \emph{Proceedings of the 33rd ACM SIGSOFT International
  Symposium on Software Testing and Analysis (ISSTA)}, 2024, pp. 1223--1235.

\bibitem{gcov}
Gcov. {A Test Coverage Program in GNU GCC tool-chain}.
  \url{https://gcc.gnu.org/onlinedocs/gcc/Gcov.html}.

\end{thebibliography}

%\newpage
%\end{thebibliography}
\newpage
\appendix

%Due to the page limit, we present extra discussions and extra experimental evaluation in the Appendix below.

\smallskip
\section*{Section A. Artifact Availability Description} \label{cottontail::appendix:artifact}

\ourSol is a new concolic execution engine guided by large language models (LLMs). 
This artifact contains the source code of
\ourSol and all benchmarks utilized in the experimental
sections of the paper. 
This document also outlines the detailed steps to
retrieve the artifact and the guidance to run the experiments, as well as how to adapt it
to real-world testing environments.

\subsection{Repository Access and Running Requirements}

\smallskip
\noindent
\textcolor{black}{(1) How to access?}
We provide public access to our code
and experiment setups through the following GitHub link:

\smallskip
\centerline{\href{https://github.com/Cottontail-Proj/cottontail}{\textcolor{blue}{https://github.com/Cottontail-Proj/cottontail}}}
\smallskip

\subsubsection{Hardware dependencies}
For a single execution of \ourSol on a subject, standard commodity machines are sufficient to meet our requirements (GPU is not required). These machines should
have a minimum of a 1-core CPU, 8GB RAM, and a 32GB hard drive. 
In this paper, all experiments were run on a Linux PC with Intel(R) Xeon(R) W-2133 CPU @ 3.60GHz x 12 processors and 64GB RAM running Ubuntu 18.04 operating system (kernel version 5.4.0).

\subsubsection{Software dependencies}
All scripts on the host system are tested on Ubuntu 18.04. However, they
are expected to work on any Linux distribution. To run these
scripts successfully, the host machines should have Python 3.9
installed along with the libraries to support GPT (e.g., {\tt openai}) and concolic execution (e.g., {\tt Z3}).
To build the customized compiler (i.e., {\tt cottontail-cc}), an LLVM compiler (LLVM-10) is needed. The tool {\it gcovr} should be installed to collect the code coverage of subject programs.

\subsubsection{Benchmarks}
All the benchmarks required for evaluation
are located within the benchmark directory of the GitHub repository below.
The folder includes the source code of the test program, building scripts, test drivers, and seed inputs to help reproduce the experiment results.

\smallskip
\centerline{\href{https://github.com/Cottontail-Proj/benchmarks}{\textcolor{blue}{https://github.com/Cottontail-Proj/benchmarks}}}
\smallskip

\subsubsection{License}
Followed by SymCC, \ourSol is free software: you can redistribute it and/or modify it under the terms of the GNU General Public License as published by the Free Software Foundation, either version 3 of the License, or (at your option) any later version.

\subsection{Artifact Installation and Configuration}

We now set up the artifact, and the entire process to finish the installation is estimated to take 2 hours.

\smallskip
\indent
\leftline{\textcolor{black}{(1) Download the artifact from GitHub:}}

\smallskip
\noindent
\newline
\centerline{%
  \parbox{0.9\columnwidth}{%
    \raggedright
    {\small \texttt{\$git clone $\backslash$ https://github.com/Cottontail-Proj/cottontail}}
  }
}
\smallskip

\smallskip
\indent
\leftline{\textcolor{black}{(2) Build the compiler pass to support concolic execution:}}
\smallskip
\newline
%\centerline{{\tt \$./build-contontail-compiler.sh}}
\centerline{%
  \parbox{0.9\columnwidth}{%
    \raggedright
    \texttt{\$./build-contontail-compiler.sh}
  }
}
\smallskip

\smallskip
\indent
\leftline{\textcolor{black}{(3) Install software dependencies}}
\smallskip
\newline
\centerline{%
  \parbox{0.9\columnwidth}{%
    \raggedright
    \texttt{\$./install-deps.sh}
  }
}
\smallskip

After the above steps, no further configuration is needed, and we
can proceed with a basic run to verify that everything is
functioning correctly.

\subsection{Running Real-world Applications}

After the installation of the runnable tool, we can now use it to test real-world software systems.

\subsubsection{Prepare the subject}

We now use the customized compiler (i.e., {\tt contontail-cc}) that supports concolic execution functionalities to build the testing subject.
\smallskip
\newline
\centerline{%
  \parbox{0.9\columnwidth}{%
    \raggedright
    \texttt{\$CC=contontail-cc ./build.sh xx}
  }
}
\smallskip

\subsubsection{Set Up the Environment Variables and Paths}
We provide a configuration file {\tt config.ini} to help easily maintain the testing of different software systems. 
Users could modify the enabled LLM model, running targets, and running settings for their own needs. The contents of the configuration file are shown as follows.

\smallskip
\centerline{%
  \parbox{0.9\columnwidth}{%
    \raggedright
    \texttt{\$vim config.ini}
  }
}
\smallskip

\begin{lstlisting}[backgroundcolor=\color{lightgray}]
[common-settings]
format = JSON
llm_model = xxx
api_key = xxx

[gcov-locations]
mainDir = xx
gcovDir = %(maindir)s/json-c/build-gcov/apps/
sourceDir = %(maindir)s/json-c/
recordFile = %(maindir)s/json-c/build-gcov/all_records.txt

[running-locations]
mainDir = xx
inputDir = %(maindir)s/input
outputDir = %(maindir)s/output
failedDir = %(maindir)s/failed-cases

[running-targets]
cottontailTarget = json_parse_cottontail
gcovTarget = json_parse_gcov

[running-params]
timeout = 43200 // running timeout
cov_timeout = 60 // interval timeout for coverage collection
\end{lstlisting}

\subsubsection{Launch the concolic testing}
Users could simply use the following command to launch the testing.
\smallskip
\newline
\centerline{%
  \parbox{0.9\columnwidth}{%
    \raggedright
    \texttt{\$python ./run-cottontail.py}
  }
}
\smallskip

\setcounter{subsection}{0} 
%\smallskip
\section*{Section B: Extra Description and Experiments} 
\label{cottontail::appendix:experiments}

\subsection{Comparison of Existing Coverage Maps} \label{cottontail::appendix:com-maps}

To perform a practical concolic execution over parsing test programs, we argue that an effective selection strategy should be structure-aware, which: %\seongmin{feel free to rollback this refactoring} The selection strategy
\begin{enumerate}[label=\#\arabic*]
    \item provides a meaningful (e.g., includes semantic information) and complete representation of program paths;
    \item records human-readable coverage information;
    \item excludes redundant structure-agnostic path constraints;
    \item has less chance of missing interesting coverage.
\end{enumerate}

There are three existing strategies to handle selection for general software, which are exhaustive search, the Bitmap \cite{SymCC-usenix20}, and the CSTG \cite{hu2024marco}, yet they do not comply with all the requirements when handling parsing test programs.

\smallskip
\noindent
{\it Exhaustive Search.}
A straightforward selection approach is to select all path constraints and try to explore all program paths. 
Such an option can satisfy requirements \#1 and \#4 but not \#2 and \#3. 
As for requirement \#3, to help understand the particular redundancy in parsing programs, taking the code snippets from Figure~\ref{cottontail::fig:mujs-parsing} as an example, parsing programs typically implement the parsing and the application (omitted for simplicity) logic structurally using {\it switch-case} statements.
Such a common practice is effective because the hidden structures in the inputs can be processed smoothly.
However, when testing such implementations using concolic testing, every input byte will go through the case statements for systematic path exploration. 
In other words, assume the size of seed input is 100, there will be 100 path constraints selected to represent only one path reflected by a case statement (e.g., Line 10 in the {\tt jsY\_lexx} function).
However, only one path constraint can bring new coverage, meaning {\it 99\% of the effort on solving the rest 99 path constraints is wasted!} 
Such an issue will be significantly amplified when the input size increases or the number of branches goes up.

\smallskip
\noindent
{\it Bitmap.}
One may try to opt for {\tt Bitmap} from \symcc \cite{SymCC-usenix20} (originally from AFL \cite{fioraldi2020afl++}), which tracks and identifies new and interesting branches during fuzzing: it maintains a fixed-size bitmap (64 KB) where each entry represents an edge in the program’s control flow graph. After each execution, AFL compares the bitmap to detect new edges, marking a test as {\it interesting} if it discovers previously untaken edges.
This helps eliminate redundant test cases by discarding inputs that do not exercise new paths, thus satisfying requirement \#3. 
However, such a strategy is neither complete due to the missing semantic information by only looking at binary level coverage, nor can it miss many interesting code coverage due to hash collisions or limited trace granularity \cite{hu2024marco,fioraldi2020afl++}, thus failing to comply with requirements \#1 and \#4.

\smallskip
\noindent
{\it CSTG.}
Another attempt is to use Concolic State Transition Graph (CSTG) from \marco \cite{hu2024marco}, which builds a global view of coverage from binary code and only selects the path constraints that are globally optimal.
Unfortunately, although it improves the selection strategy by prioritizing the path constraints to solve from a global view, it shares the same drawbacks as {\tt Bitmap} for requirements \#1 and \#4. 
This is because CSTG is built upon the coverage from binaries, and the prioritized selection can miss many interesting coverage due to the lack of local optimal cases when performing global selection. 
Regarding requirement \#2, to our knowledge, no prior coverage map could efficiently be used to provide human-readable coverage information at runtime. 
Not that it is possible to retrieve the coverage information from external tools (e.g., {\tt gcov} \cite{gcov}), but it could be time-consuming to get the coverage results.

\begin{comment}
\subsection{Response Examples of Seed Generation Prompts}

\begin{figure}[t] 
  \begin{lstlisting}[escapechar=@,frame=single,numbers=none] 
  //Format: fileName_funcName_lineNumber_columnNumber_brType_brId
  
  //Example of covered branches:
  quickjs.c_js_parse_expr_binary_24289_13_switch_37
  quickjs.c_js_parse_stmt_22890_8_if_1
  
  //Example of uncovered branches:
  quickjs.c_js_parse_expr_binary_24289_13_switch_42
  quickjs.c_js_parse_stmt_22890_8_if_0
  \end{lstlisting} 
  \vspace{-1em}
  \caption{Branch coverage information (covered/uncovered) recorded}
  \label{cottontail::fig:branch-record} 
  \end{figure}

\begin{figure}[t]
  \centering
  %\epsfig{file=image/fm.bmp, width=8.5cm}
  \includegraphics[width=0.99\linewidth]{pictures/seed-prompt.pdf}
  \vspace{-1em}
  \caption{Response of LLM-driven Seed Generation (from Option 1) \todo{x}}
  \label{cottontail::fig:prompt-seed-generation-response1}
  \vspace{-1em}
\end{figure}

\begin{figure}[t]
  \centering
  %\epsfig{file=image/fm.bmp, width=8.5cm}
  \includegraphics[width=0.99\linewidth]{pictures/seed-prompt.pdf}
  \vspace{-1em}
  \caption{Response of LLM-driven Seed Generation (from Option 2) \todo{x}}
  \label{cottontail::fig:prompt-seed-generation-response2}
  \vspace{-1em}
\end{figure}
\end{comment}

\begin{table}[t]
	\centering
    \scriptsize
	%\vspace{-1mm}
	\caption{Comparison of different coverage maps implemented in current concolic execution engines}
        %\vspace{-1em}
	%\begin{tabular*}{\hsize}{@{}@{\extracolsep{\fill}}ccccc@{}}
	\begin{tabular}{ccccc}
		\toprule
        \multirow{2}*{\textbf{Req}} & 
		\multirow{1}*{\textbf{\symccNoMap}} & 
        \multicolumn{1}{c}{\textbf{\symcc}} &
		\multicolumn{1}{c}{\textbf{\marco}} &
		\multicolumn{1}{c}{\textbf{\ourSol}} \\
		&   \textit{(w/o bitmap)} & \textit{(w/ bitmap)} & {\it (w/ CTSG)} & {\it (w/ ECT)}\\
		\midrule
        \multirow{1}{*}{\#1} & \ding{52}  & \ding{56} & \ding{56} & \ding{52}  \\
		\multirow{1}{*}{\#2} & \ding{56}  & \ding{56} & \ding{56} & \ding{52}  \\
		%\multirow{1}{*}{Eff. {\it pc} solving} & Low  & Low & High & High  \\
		\multirow{1}{*}{\#3} & \ding{56}  & \ding{52} & \ding{52} & \ding{52}  \\
        \multirow{1}{*}{\#4} & \ding{52}  & \ding{56} & \ding{56} & \ding{52}  \\
		
		\bottomrule
	\end{tabular} 
	%\vspace{-2em}
	\label{cottontail::tab:approach-comparison-maps}
\end{table}

\begin{table*}[t]
	\centering
	\caption{Line and branch coverage comparison results with existing concolic execution engines \symcc \cite{SymCC-usenix20} and \marco \cite{hu2024marco}} 
	\vspace{-1em}
	\begin{tabular}{c} 
	    \includegraphics[width=0.99\linewidth]{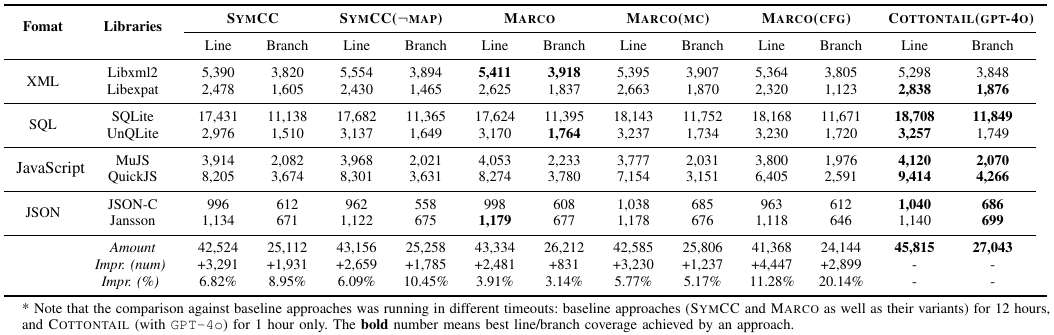}
    \end{tabular}
   \label{cottontail::tab:evaluation:1v12}
    \vspace{-1em}
\end{table*}
\subsection{Extra Experimental Results against State-of-the-art Approaches} \label{cottontail::appendix:extra-experiments}

%\smallskip
\noindent
{\bf Results of Running Setting-2.}
Table \ref{cottontail::tab:evaluation:1v12} presents the line and branch coverage results of \ourSol against the baseline approaches \symcc, \marco, and their variants.
As shown in Table \ref{cottontail::tab:evaluation:1v12}, \ourSol consistently achieves the highest line and branch coverage in the majority of the evaluated benchmarks, although it was allocated only one hour of execution time - significantly less than the 12-hour time budget used by baseline approaches such as \symcc and \marco (and their respective variants). In particular, \ourSol attains the best line coverage in 7 of 8 programs. On average, it covers up to 4.4k lines (6.77\% on average) more and 2.8k branches (9.57\% on average) more in total than all other techniques. These results highlight the effectiveness and efficiency of \ourSol in rapidly exploring diverse program paths and uncovering deep execution behaviors, even under constrained time settings.
There are a few cases where \ourSol covers less branch coverage than \marco (e.g., in Libxml2), this is because \marco designs the path selection based on random sampling, so a few more line/branch coverages are expected.

\begin{figure*}[t]
	\centering
	\includegraphics[width=1\linewidth]{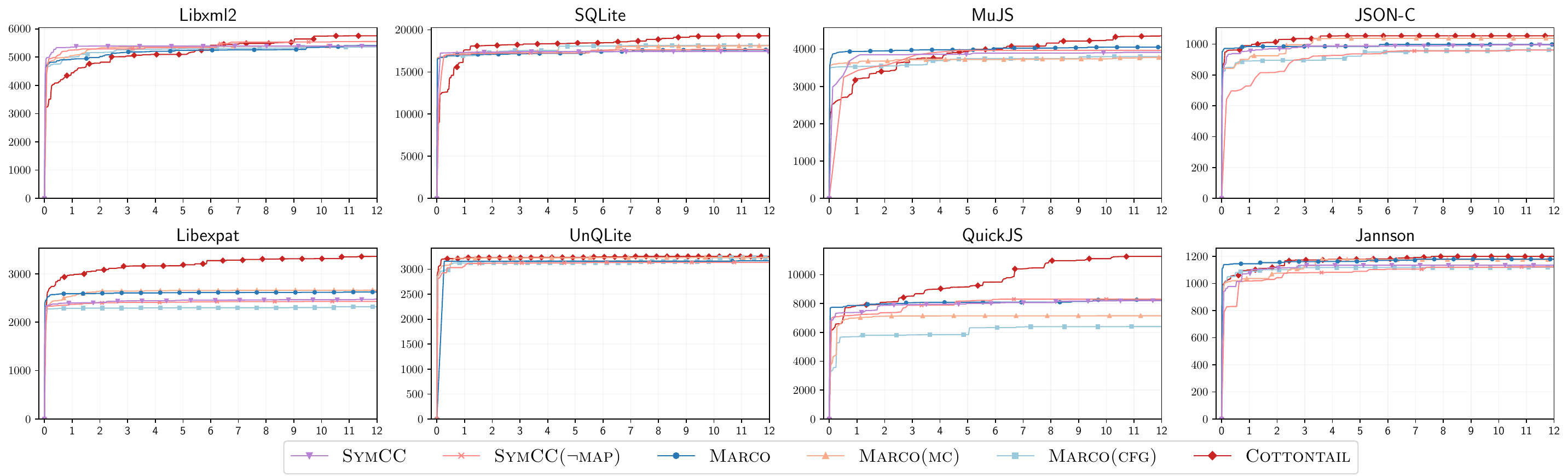}
	\vspace{-1em}
	\caption{Line coverage comparison among \ourSol and baseline approaches in 12 hours ({\it x-axis} indicates line coverage while {\it y-axis} the time)}
	\label{cottontail::fig:rq1-12h}
	\vspace{-1em}
\end{figure*}

\smallskip
\noindent
{\bf Results of Running Setting-3.}
Figure \ref{cottontail::fig:rq1-12h} illustrates the code coverage trend over time. We can see that \ourSol performs better than {\it all} comparative approaches within 12 hours, achieving higher line coverage rates from 15.10\% to 21.41\% on average.
Note that we changed the base model from {\tt GPT-4o} to {\tt GPT-4o-mini} in this setting, as {\tt GPT-4o} is already superior to baseline approaches in one hour, and {\tt GPT-4o-mini} is more cost-effective.

It is worth noting that on many benchmarks (e.g., Libexpat and SQLite), both \symcc and \marco exhibit early saturation in their coverage progress. 
For example, \symcc quickly reaches a plateau within one hour and shows minimal improvement thereafter, indicating limited capability in uncovering additional program behaviors beyond its initial exploration. 
The final coverage achieved by both techniques remains substantially lower than that of other approaches, suggesting that their underlying strategies are less effective in sustaining exploration over time.
Thus, fresh seeds are required to change the saturation and make the testing more effective, which motivates us to design a new seed generation strategy in \ourSol to help explore more program paths. The contribution of the seed acquisition is presented in \S \ref{cottontail::sec:approach:seed-generation}.

\subsection{Overhead of Structural Instrumentation} \label{cottontail::eva:rq2.1-overhead}

Since performance is important when conducting concolic execution, it would be better to know how much overhead is introduced by structural instrumentation. %(presented in \S \ref{cottontail::sec:approach:instrumentation}).
We then compared the compilation time between SymCC (without structural instrumentation) and \ourSol.

Table \ref{cottontail::tab:evaluation:rq2.1} presents the overhead in seconds introduced by structural instrumentation for various libraries when using \symcc and \ourSol.
The overhead ranges from low to moderate across the libraries, with JSON-C and Jansson exhibiting the least overhead (9.78\% and 23.96\%, respectively), while Libexpat and QuickJS experience higher overheads (31.49\% and 26.26\%, respectively). 
%The number of structural instrumentation (SI) listed indicates that more complex libraries tend to result in slightly higher overhead.

\begin{table}[t]
\centering
    %\scriptsize
	%\vspace{-1mm}
	\caption{Overhead (second) of structural instrumentation}
        %\vspace{-1em}
	%\begin{tabular*}{\hsize}{@{}@{\extracolsep{\fill}}ccccc@{}}
	\begin{tabular}{cccc}
		\toprule
        \multirow{1}*{\textbf{Libraries}} & 
		\multirow{1}*{\textbf{SymCC}} & 
        \multicolumn{1}{c}{\textbf{\ourSol}} &
		\multicolumn{1}{c}{\textbf{Overhead}} \\
		\midrule
       
		\multirow{1}{*}{Libxml2} &  17.61  & 21.20  & 20.39\%  \\
		\multirow{1}{*}{Libexpat} &  6.86  & 9.02 & 31.49\%  \\
		%\multirow{1}{*}{curl} 	& 13.93  & 15.35 (30)  & 10.19\%  \\
		%\multirow{1}{*}{lexbor} & 77.40 & 86.56 (338)  & 11.83\%  \\
		\multirow{1}{*}{SQLite} & 73.78  &  81.27 & 10.15\%  \\
		\multirow{1}{*}{UnQLite} & 13.14  & 14.99  & 14.08\%  \\
		\multirow{1}{*}{MuJS}    & 11.45  & 13.37   & 16.77\%  \\
		\multirow{1}{*}{QuickJS} & 306.11  & 386.50  & 26.26\%  \\
            \multirow{1}{*}{JSON-C} & 3.68  & 4.04 & 9.78\%  \\
		\multirow{1}{*}{Jansson} & 2.88  & 3.57 & 23.96\%  \\
		\bottomrule
	\end{tabular}
	%\vspace{-1em}
	\label{cottontail::tab:evaluation:rq2.1}
\end{table}

It is worth noting that despite the relative increase in execution time due to instrumentation, the overhead remains negotiable when compared to the significantly longer durations required for comprehensive testing. 
Structural instrumentation provides critical insights into program execution and ensures higher code coverage, making a better trade-off in performance and effectiveness in exhaustive testing.

\subsection{Performances of Integration with Other LLMs} \label{cottontail::appendix:different-llms}

%\subsubsection{Integration with {\tt GPT-4o}}
%To investigate the performance of \ourSol with more intelligent (costly) LLMs, we conducted experiments using the {\tt GPT-4o}\footnote{\url{https://platform.openai.com/docs/models/gpt-4o}} model, which is a smaller version of the original {\tt GPT-4}\footnote{\url{https://platform.openai.com/docs/models/gpt-4}} model.

%\subsection{Different Impact of Parameter Selection for Equation \ref{cottontail::eq:selector}}

\begin{figure*}[t]
	\centering
	\includegraphics[width=1\linewidth]{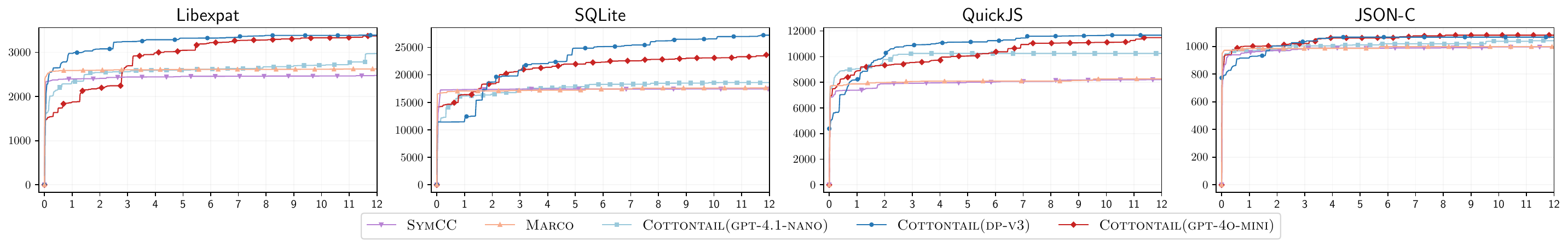}
	\vspace{-1em}
	\caption{Line coverage comparison among variants that integrate different LLMs in 12 hours: \ourSol uses {\tt gpt-4o-mini} as base model, so \ourSolmini and \ourSol are identical. ({\it x-axis} indicates line coverage while {\it y-axis} the time)}
	\label{cottontail::fig:evaluation:deepseek}
	\vspace{-1em}
\end{figure*}

%\subsubsection{Integration with {\tt DeepSeek-V3} and {\tt GPT-4.1-nano}}
To investigate the impact of different LLMs on the performance of \ourSol, we conducted experiments using two additional recently released LLMs: open-source {\tt DeepSeek-V3} released on 20/01/2025 and {\tt GPT-4.1-nano} released on 14/04/2025.
\begin{itemize}[leftmargin=1em,nosep]
  \item \ourSoldp: the version of \ourSol that integrates with {\tt DeepSeek-V3}\footnote{\url{https://github.com/deepseek-ai/DeepSeek-V3}} as the base model.
  \item \ourSolnano: the version of \ourSol that integrates with {\tt GPT-4.1-nano}\footnote{\url{https://platform.openai.com/docs/models/gpt-4.1-nano}} model.
  \item \ourSolmini: the default version of \ourSol that integrates {\tt GPT-4o-mini} as base model.
\end{itemize}
We run the comparative experiments on the same set of benchmarks as in the previous section in \S\ref{cottontail::eva:rq2.4-seed-generation}, with a time budget of 12 hours for each LLM.
The results are shown in Figure \ref{cottontail::fig:evaluation:deepseek}.
The results demonstrate that \ourSol is highly adaptable and effective when integrated with various open-source and closed-source LLMs. 
As shown in the figures, \ourSol achieves competitive or superior performance across different LLMs, including {\tt GPT-4.1-nano} and {\tt DeepSeek-V3}. Notably, \ourSoldp achieves the best line coverage in most cases, while the performance with other LLMs remains consistently strong. This highlights the flexibility of our framework to leverage the strengths of different LLMs, whether open-source or proprietary, ensuring robust and efficient concolic execution across diverse testing environments.
Furthermore, the integration cost is minimal, making it feasible to incorporate \ourSol into recent advanced LLMs without significant overhead. This low-cost adaptability ensures that \ourSol can seamlessly enhance the testing capabilities of modern LLM-driven systems.

\newpage
\section*{Section C: Meta-Review}

The following meta-review was prepared by the program
committee for the 2026 IEEE Symposium on Security and
Privacy (S\&P) as part of the review process as detailed in the call for papers.

\subsection*{Summary of Paper}

This paper presents \ourSol, a novel concolic execution framework to overcome the several key limitations (i.e., structure-agnostic constraint selection, syntax-ignorant solving, and reliance on manual or random seed inputs) in the existing concolic execution systems, by leveraging Large Language Models (LLMs).

\subsection*{Scientific Contributions}

\begin{itemize}[leftmargin=1em,nosep]
  \item Creates a New Tool to Enable Future Science
  \item Addresses a Long Known Issue
  \item Provides a Valuable Step Forward in an Established Field
\end{itemize}

\subsection*{Reasons for Acceptance}

\begin{itemize}[leftmargin=1em,nosep]
  \item 1. The use of LLMs as integral components in concolic execution is innovative.
  \item 2. The paper tackles several long-standing issues in concolic execution, and the experiments show that the proposed solution indeed addresses these issues effectively.
  \item 3. The prototype \ourSol is open-sourced to enable future science. It not only allows further improvement on concolic execution, but also other software security research that uses concolic execution as an analysis tool.
\end{itemize}

\subsection*{Noteworthy Concerns}

\begin{itemize}[leftmargin=1em,nosep]
  \item 1. The paper's evaluation targets are limited to small programs that rely on structured input formats, which are well understood by LLMs. Its performance on large programs, programs with unstructured inputs, and programs with input formats unknown by LLMs is not evaluated.
  \item 2. While the Discussion and Related Work carefully place the proposed technique into a broader context (i.e., automated testing techniques, including fuzzing), such contextualization comes late in the paper -- those who only read the introduction and evaluation may fail to appreciate the technique's limitations.
\end{itemize}

%\input{sections/Appendix-rebuttal}

\end{document}